\documentclass[12pt, a4paper]{article}

\usepackage{url}
\usepackage{amsmath}
\usepackage{amssymb, parskip, listings, mathrsfs, mathtools} 
\usepackage{graphicx}
\usepackage[absolute,overlay]{textpos}
\usepackage{helvet}
\usepackage{array}
\usepackage{color}
\usepackage{setspace}
\usepackage{booktabs}
\usepackage{changepage}
\newcommand{\colvec}[2][0.9]{%
  \scalebox{#1}{%
    \renewcommand{\arraystretch}{0.9}%
    $\begin{bmatrix}#2\end{bmatrix}$%
  }
}
\usepackage{tikz}
\usetikzlibrary{positioning}

\usepackage{geometry, graphicx, amsmath, amssymb, amsthm, parskip, listings, color} 
\usepackage[font=small,labelfont=bf]{caption}
\usepackage{natbib}

\usepackage[affil-it]{authblk} 
\usepackage{etoolbox}
\usepackage{lmodern}



\newcommand*{\affaddr}[1]{#1} 
\newcommand*{\affmark}[1][*]{\textsuperscript{#1}}

\title{\textbf{Bayesian Hierarchical Models for High-Dimensional Mediation Analysis with Coordinated Selection of Correlated Mediators}}

\author{
\textbf{Yanyi Song}$^{1}$, \textbf{Xiang Zhou}$^{1,*}$, \textbf{Jian Kang}$^{1,*}$, \textbf{Max T. Aung}$^{1}$, \textbf{Min Zhang}$^{1}$, \textbf{Wei Zhao}$^{2}$, \\ \vspace{-0.1in} \textbf{Belinda L. Needham}$^{2}$, \textbf{Sharon L. R. Kardia}$^{2}$, \textbf{Yongmei Liu}$^{3}$, \textbf{John D. Meeker}$^{4}$,  \textbf{Jennifer A. Smith}$^{2}$, \textbf{and Bhramar Mukherjee}$^{1}$ \\
\affaddr{\affmark[1]Department of Biostatistics, School of Public Health, University of Michigan, Ann Arbor, MI, U.S.A.} \\
\affaddr{\affmark[2]Department of Epidemiology, School of Public Health, University of Michigan, Ann Arbor, MI, U.S.A.}\\
\affaddr{\affmark[3]Department of Medicine, Divisions of Cardiology and Neurology, Duke University, Durham, NC, U.S.A.} \\
\affaddr{\affmark[4]Department of Environmental Health Sciences, School of Public Health, University of Michigan, Ann Arbor, MI, U.S.A.} \\
\affaddr{\affmark[$*$]\textit{email}:\textnormal{ xzhousph@umich.edu}} \\
\affaddr{\affmark[$**$]\textit{email}:\textnormal{ jiankang@umich.edu}}
}

\begin{document}
\date{}
\maketitle

\begin{abstract}
We consider Bayesian high-dimensional mediation analysis to identify among a large set of correlated potential mediators the active ones that mediate the effect from an exposure variable to an outcome of interest. Correlations among mediators are commonly observed in modern data analysis; examples include the activated voxels within connected regions in brain image data, regulatory signals driven by gene networks in genome data and correlated exposure data from the same source. When correlations are present among active mediators, mediation analysis that fails to account for such correlation can be sub-optimal and may lead to a loss of power in identifying active mediators. Building upon a recent high-dimensional mediation analysis framework, we propose two Bayesian hierarchical models, one with a Gaussian mixture prior that enables correlated mediator selection and the other with a Potts mixture prior that accounts for the correlation among active mediators in mediation analysis. We develop efficient sampling algorithms for both methods. Various simulations demonstrate that our methods enable effective identification of correlated active mediators, which could be missed by using existing methods that assume prior independence among active mediators. The proposed methods are applied to the LIFECODES birth cohort and the Multi-Ethnic Study of Atherosclerosis (MESA) and identified new active mediators with important biological implications.
\end{abstract}

\section{Introduction}
\label{sec1}
Mediation analysis attempts to explain the intermediate mechanism through which an exposure affects an outcome, and quantify the indirect effect transmitted by the mediator variable between the exposure and the outcome \citep{mackinnon2008introduction}. To formally define the direct and indirect effects, a causal approach to mediation analysis based on the counterfactual framework has been proposed, with the key assumptions for identification and causal interpretation being specified \citep{imai2010general, pearl2012causal}. This framework further gave rise to other extensions in mediation analysis, such as exposure-mediator interaction \citep{valeri2013mediation}, survival data \citep{vanderweele2011causal}, etc. 

The fast development in high-throughput biological technology has provided tremendous opportunities for mediation analysis with large-scale omics data. Modern omics studies often collect a large number of mediators with the goal for identifying active mediators that mediate the effect from an exposure variable to an outcome variable. In many of these modern data applications, there often exists a substantial correlation among mediators. For example, in functional MRI (fMRI) studies, the brain images are composed of a large number of voxels/regions and true signals usually represent connected regions. Our study is particularly motivated by two large-scale data, one in environmental science and one in genomics. The first is the LIFECODES birth cohort, one of the nation’s largest pregnancy cohorts aimed at advancing care and improving outcomes in high-risk pregnancies \citep{mcelrath2012longitudinal}. This study collected data on a large group of endogenous biomarkers of lipid metabolism, inflammation, and oxidative stress. These biomarkers are hypothesized to mediate the effects of prenatal exposure to environmental contamination on adverse pregnancy outcomes \citep{Aung2020.05.30.20117655}. Moderate to strong correlations across those biomarkers are observed, and such correlations occur not only for biomarkers within the same biological pathways but also for biomarkers between different pathways. The second is the Multi-Ethnic Study of Atherosclerosis (MESA) data \citep{bild2002multi}. In this study, high-dimensional DNA methylation (DNAm) are hypothesized to mediate the effect of neighborhood factors on blood glucose level, which is a critical variable linked to diabetes and heart diseases. Like the first study, these DNAm data are also correlated with each other. Performing mediation analysis with a high-dimensional set of mediators that may be correlated with each other is an important first step towards understanding the molecular basis of complex diseases and subsequent development of prevention and treatment strategies.

Several mediation analysis methods have been recently developed to accommodate high-dimensional mediators obtained from large-scale genomic data. For example, \cite{zhang2016estimating} proposes sure independent screening and minimax concave penalty techniques to study how the high-dimensional DNAm mediate the effect of smoking on lung function; \cite{zhao2016pathway} develops a new convex, Lasso-type penalty on the indirect effects to identify brain pathways from the language stimuli to the outcome region activity. In addition to the frequentist methods, \cite{song2018bayesian} proposes a Bayesian variable selection method with separate shrinkage priors on the exposure-mediator effects and mediator-outcome effects, respectively. \cite{song2020mediation} further replaces the two separate priors with relevant joint priors for a direct target on the non-zero indirect effect in mediator selection. Those methods enable a joint analysis of high-dimensional mediators and a valid procedure for the identification of active mediators. However, to the best of our knowledge, none of the existing methods for high-dimensional mediation analysis has accounted for the possible correlation structure among active mediators. As explained in the above paragraph, such correlation is highly prevalent. When the truly active mediators are correlated with one another, then the existing methods that fail to account for such correlation may lead to a loss of power. A more effective mediation analysis will require methods that can incorporate the useful correlation information of high-dimensional mediators into the model building process. We attempt to fill this gap in the literature.

Our proposed methods are based on a recently developed high-dimensional mediation analysis framework \citep{song2020mediation}, which introduced a Gaussian mixture model (GMM) as a joint prior on the exposure-mediator and mediator-outcome effect to allow for a targeted penalization on the indirect effect. This method has been shown to enjoy excellent and robust performance for mediator selection and effect estimation. GMM assumes that each mediator can be independently categorized into one of the four components based on association pattern, and its group indicator follows the same multinomial distribution as the other mediators. With the goal of utilizing the correlation structure among mediators in the modeling process, we aim to replace the independent priors on the mediators' group indicators with two priors that introduce coordinated selection on active mediators that may be correlated with each other. One prior is based on the Potts distribution  \citep{potts1952some}, a generalization from the Ising distribution, which allows for more than two groups and complex dependency between correlated neighboring variables. The other prior is based on a jointly modeling of the mediator-specific mixing probabilities via a logistic normal distribution \citep{atchison1980logistic}, with the group probabilities reflecting the underlying correlation structure. Both methods allow for high-dimensional mediation analysis with the possible coordinated selection of active mediators via another layer in the Bayesian hierarchy. Both methods are built off the GMM proposed in \cite{song2020mediation}, and thus inherit the merits of the GMM method for high-dimensional mediation analysis. Furthermore, the proposed methods incorporate the structural information into a prior that favors selection of correlated mediators, and are expected to allow the identification of correlated active mediators that could be missed otherwise. Our methods rely on exact posterior sampling to provide estimates of quantities of interest and characterize uncertainty in estimation. The proposed methods will also facilitate the interpretation of the results, particularly for the selected mediators with high correlations.

We note that our methods are built upon a long history of similar methods in other related statistics areas. Indeed, Bayesian variable selection with covariate structural information has received much interest over the years. Bayesian group Lasso \citep{raman2009bayesian} and Bayesian sparse group selection method \citep{chen2016bayesian} allow for the inclusion of grouping effects and lead to more parsimonious models with reduced estimation error compared with standard Lasso. \cite{yuan2005efficient} also develop a correlation prior on the binary selection indicators to distinguish models with the same size. Bayesian graphical models represent another stream of work on structural variable selection. \cite{cai2018bayesian} utilizes the graph Laplacian matrix to encode the network information into the regression coefficients. \cite{stingo2011incorporating} proposes the simultaneous selection of pathways and genes, using the pathway summaries of the group behavior and structure dependency within pathways to inform the selection. Along with the above methods, emerging literature considers the extension of the ``spike-and-slab" type of mixture prior \citep{mitchell1988bayesian} in combination with Markov random field (MRF) prior to incorporate graph information. Ising prior, a binary spatial MRF, and its variations have been effectively applied to induce sparsity and accommodate selection dependency. \cite{li2010bayesian} and \cite{chekouo2016bayesian} show that the structural information through Ising priors can greatly improve selection and prediction accuracy over the independent priors. In addition to smoothing over the latent selection indicators, recent studies deploy different types of ``slab distribution", such as the Dirichlet Process \citep{li2015spatial}, the group fused Lasso prior \citep{zhang2014bayesian}, etc., to include the grouping and smoothing effect in the non-zero regression coefficients due to local dependence or high correlation. Those methodologies have illustrated how the structural or correlated information can be incorporated into Bayesian framework to deliver better variable selection. However, these existing approaches are not designed specifically for mediation models with multivariate mediators and thus not directly applied to high-dimensional mediation analysis. 

The rest of the paper is organized as follows. In Section \ref{sec2}, we first define the causal effects of interest for the multivariate mediation analysis with the counterfactual framework. Then we review the mediation estimands under the linear regression models with multiple mediators and one continuous outcome. 
In Section \ref{sec3}, we propose two novel methods to explicitly incorporate correlation structure among mediators while jointly analyzing them. Simulation studies are carried out and discussed in Section \ref{sec4}. We illustrate our methods by applying them to LIFECODES and MESA cohort in Section \ref{sec5}, and conclude the paper with a discussion in Section \ref{sec6}.

\section{Notations, Definitions and Models}
\label{sec2}
We adopt the counterfactual framework for causal mediation analysis in a high-dimensional setting. Consider a study of $n$ subjects and for subject $i$, $i=1,\ldots, n$, we collect data on one exposure $A_i$, $p$ potential mediators $\boldsymbol{M}_i=(M_i^{(1)},M_i^{(2)}, \ldots, M_i^{(p)})^\top$, one outcome $Y_i$, and $q$ covariates $\boldsymbol{C}_i=(C_i^{(1)}, \ldots, C_i^{(q)})^\top$. In particular, we focus on the case where $Y_i$ and $\boldsymbol{M}_i$ are all continuous variables. We define $\boldsymbol{M}_i(a) = (M_i^{(1)}(a),M_i^{(2)}(a), \ldots, M_i^{(p)}(a) )$ as the $i$th subject's counterfactual value of the $p$ mediators if he/she received exposure $a$, and define $Y_i(a,\boldsymbol{m})$ as the $i$th subject's counterfactual outcome if the subject's exposure were set to $a$ and mediators were set to $\boldsymbol{m}$. The effect of an exposure can be decomposed into its direct effect and effect mediated through mediators, i.e. indirect effect. The natural direct effect (NDE) of the given subject is defined as $Y_i(a, \boldsymbol{M}_i(a^{\star})) - Y_i(a^{\star}, \boldsymbol{M}_i(a^{\star}))$, where the exposure changes from $a^{\star}$ (the reference level) to $a$ and mediators are hypothetically controlled at the level that would have naturally been with exposure $a^{\star}$. The natural indirect effect (NIE) of the given subject is defined by $Y_i(a, \boldsymbol{M}_i(a)) - Y_i(a, \boldsymbol{M}_i(a^{\star}))$, the change in counterfactual outcomes when mediators change from  $\boldsymbol{M}_i(a^{\star})$ to $\boldsymbol{M}_i(a)$ while fixing exposure at $a$. The total effect (TE), $Y_i(a, \boldsymbol{M}_i(a)) - Y_i(a^{\star}, \boldsymbol{M}_i(a^{\star}))$, can then be expressed as the summation of the NDE and the NIE: $Y_i(a, \boldsymbol{M}_i(a)) - Y_i(a^{\star}, \boldsymbol{M}_i(a^{\star})) =Y_i(a, \boldsymbol{M}_i(a)) - Y_i(a, \boldsymbol{M}_i(a^{\star}))+ Y_i(a, \boldsymbol{M}_i(a^{\star})) - Y_i(a^{\star}, \boldsymbol{M}_i(a^{\star}))  = $ NIE + NDE. 

The counterfactual variables are useful concepts to formally define causal effects, but they are not necessarily observed. In order to estimate the average NDE and NIE from observed data, further assumptions are required, including the consistency assumption and four non-unmeasured confounding assumptions (\citealp{vanderweele2016mediation}). We elaborate those assumptions in Section 1 of the Supplementary Materials (SM). It has been shown that under those assumptions, the average NDE and NIE can be identified by modeling $Y_i | A_i, \boldsymbol{M}_i, \boldsymbol{C}_i$ and $\boldsymbol{M}_i | A_i, \boldsymbol{C}_i$ using observed data (\citealp{song2018bayesian}). Therefore, we can work with the two conditional models for $Y_i | A_i, \boldsymbol{M}_i, \boldsymbol{C}_i$ and $\boldsymbol{M}_i | A_i, \boldsymbol{C}_i$, and subsequently propose two linear models for these two conditional relationships. For the outcome model, we assume
\begin{equation}
Y_i = \boldsymbol{M}_i^\top\boldsymbol{\beta_m} + A_i\beta_a + \boldsymbol{C}_i^\top\boldsymbol{\beta_c} + \epsilon_{Yi},
\label{eq:outcome}
\end{equation}
where $\boldsymbol{\beta_m} = (\beta_{m1}, \ldots, \beta_{mp})^\top$; $\boldsymbol{\beta_c} = (\beta_{c1}, \ldots, \beta_{cq})^\top$; and $\epsilon_{Yi} \sim \textnormal{N}(0, \sigma_e^2)$. For the mediator model,  we consider a multivariate regression model that jointly analyzes all $p$ potential mediators together as dependent variables:
\begin{equation}
\boldsymbol{M}_i = A_i\boldsymbol{\alpha_a} + \boldsymbol{\alpha_c}\boldsymbol{C}_i + \boldsymbol{\epsilon}_{Mi},
\label{eq:mediator}
\end{equation}
where $\boldsymbol{\alpha_a} = (\alpha_{a1}, \ldots, \alpha_{ap})^\top$; $\boldsymbol{\alpha_c} = (\boldsymbol{\alpha^\top_{c1}}, \ldots, \boldsymbol{\alpha^\top_{cp}})^\top$, $\boldsymbol{\alpha_{c1}}, \ldots, \boldsymbol{\alpha_{cp}}$ are $q$-by-1 vectors; $\boldsymbol{\epsilon}_{Mi} \sim \mathrm{MVN}(\boldsymbol{0}, \boldsymbol{\Sigma})$, with $\boldsymbol{\Sigma}$ capturing the residual error covariance. $\epsilon_{Yi}$ and $\boldsymbol{\epsilon}_{Mi}$ are assumed to be independent of each other and independent of $A_i$ and $\boldsymbol{C}_i$. Under the identifiability assumptions discussed in SM and the modeling assumptions (linearity, no exposure-mediator interaction in the outcome and mediator model) in (\ref{eq:outcome})-(\ref{eq:mediator}), we can express causal effects with the model coefficients as below (\citealp{song2018bayesian}). In the rest of the paper, we refer to NDE as direct effect and NIE as indirect/mediation effect.
\begin{eqnarray*}
\quad \textnormal{NDE} &=& E[Y_i(a, \boldsymbol{M}_i(a^{\star})) - Y_i(a^{\star}, \boldsymbol{M}_i(a^{\star}))|\boldsymbol{C}_i]=\beta_a(a - a^\star). \\
\quad \textnormal{NIE} &=& E[Y_i(a, \boldsymbol{M}_i(a)) - Y_i(a, \boldsymbol{M}_i(a^{\star}))|\boldsymbol{C}_i] = (a - a^\star)\boldsymbol{\alpha^\top_a}\boldsymbol{\beta_m} = (a - a^\star)\sum_{j=1}^p \alpha_{aj}\beta_{mj}. \label{eq:sumIE}\\
\quad \textnormal{TE} &=& E[Y_i(a, \boldsymbol{M}_i(a)) - Y_i(a^{\star}, \boldsymbol{M}_i(a^{\star}))|\boldsymbol{C}_i] = (\beta_a + \boldsymbol{\alpha^\top_a}\boldsymbol{\beta_m})(a - a^\star).
\end{eqnarray*}

\section{Method}
\label{sec3}
Recent application of univariate mediation analysis methods at genome-wide scale (\citealp{huang2019genome, huang2019variance}) recognize the need for decomposing the null hypothesis of zero indirect effect into three null components: zero exposure on mediator effect; zero mediator on outcome effect; and both. Such composite structure of the null hypothesis in the univaraite mediation analysis can be naturally captured by the four-component Gaussian mixture model developed in the presence of high-dimensional mediators \citep{song2020mediation}. Following \citealp{song2020mediation}, we also consider a four-component Gaussian mixture for the effects of the $j$-th mediator,
\begin{eqnarray*}
[ 
\beta_{mj}, \alpha_{aj}  ]^\top \sim \pi_{1j}\mathrm{MVN}_2(\boldsymbol{0}, \boldsymbol{V}_1) + \pi_{2j}\mathrm{MVN}_2(\boldsymbol{0}, \boldsymbol{V}_2) + \pi_{3j}\mathrm{MVN}_2(\boldsymbol{0}, \boldsymbol{V}_3) + \pi_{4j}\boldsymbol{\delta}_0
\end{eqnarray*}
with a prior probabilities $\pi_{kj}$ ($k \in \Omega, \Omega = \{ 1, 2, 3, 4 \} $) summing to one and $\mathrm{MVN}_2$ denoting a bivariate Gaussian distribution. The first component represents active mediators, where both the exposure-mediator effect $\alpha_{aj}$ and mediator-outcome effect $\beta_{mj}$ are non-zero and $\boldsymbol{V}_1$ models their covariance. The inactive mediator will fall into one of the remaining three components. The second component corresponds to mediators with non-zero $\beta_{mj}$ but zero $\alpha_{aj}$, and the third component corresponds to mediators with non-zero $\alpha_{aj}$ but zero $\beta_{mj}$. Both $\boldsymbol{V}_2$ and $\boldsymbol{V}_3$ are low-rank matrices restricting that only $\beta_{mj}$ or $\alpha_{aj}$ is non-zero. Mediators with both exposure-mediator effect and mediator-outcome effect being zero belong to the fourth component, and $\boldsymbol{\delta}_0$ is a point mass at zero.

We introduce a membership indicator variable $\gamma_j$ for the $j$-th mediator, where $\gamma_j = k$ if $\large [
\beta_{mj}, \alpha_{aj} \large ]^{\top}$ is from Gaussian component $k$, $k \in \{ 1, 2, 3, 4 \} $. If we assume independence among $\pi_{k1}, \pi_{k2}, \ldots, \pi_{kp}$ (and subsequently $\gamma_1, \gamma_2, \ldots, \gamma_p$), then each mediator is independent \textit{a priori} and the prior distribution on $ [
\boldsymbol{\beta_m}, \boldsymbol{\alpha_a} ]^{\top}$ after integrating out $\{ \pi_{kj} \}$ (or $\{ \gamma_j \}$) is essentially a separable product of distributions of $\large [
\beta_{mj}, \alpha_{aj} \large ]^{\top}$. This is akin to the concept of ``separable prior" in \cite{rovckova2018spike}. In contrast, the previously developed GMM method \citep{song2020mediation} assumes a common set of $\pi_{1}, \pi_{2}, \pi_{3}, \pi_{4}$ for all the mediators {\it a priori}. This specification ties mediators together through the mixing probabilities and enables information sharing across mediators, making the priors ``non-separable". However, since this previous GMM approach assumes the same mixing probabilities for all the mediators \textit{a priori}, it does not differentiate highly correlated mediators from uncorrelated ones to inform coordinated mediator selection. Specifically, when the $j$-th and $(j+1)$-th mediators are highly correlated with each other, because such correlation often implies common biological mechanism underlying both mediators, then one mediator being active becomes informative on the other being active in the sense that $\gamma_j$ and $\gamma_{j+1}$ are more likely to be same. To enable coordinated selection of correlated active mediators, we consider embedding the correlation information to $\{ \pi_{kj} \}$'s or $\gamma_j$'s. In the following sections, we describe the proposed methods with more details.

\subsection{Hierarchical Potts Mixture Model: GMM-Potts}
\label{sec31}
The Potts model \citep{potts1952some} was initially developed as a generalization of the Ising model in statistical physics. However, it has enjoyed great success as a prior model for the spatial modeling in image analysis \citep{feng2012mri,li2019bayesian}, disease mapping \citep{best2005comparison}, genetics studies \citep{yu2012flexible}, etc. In those applications, Potts models incorporate spatial Markovian dependency by assigning homogeneous relationships for the ``neighboring"  regions. In the context of mediation analysis, we allocate the high-dimensional mediators into four Gaussian components based on their exposure-mediator and mediator-outcome effects. We think of the highly correlated mediators as neighbors and we attempt to assign them to different mediation components through a Potts model. 

To specifically formulate our Potts mixture model, we assume that $\boldsymbol{\gamma} = (\gamma_1, \gamma_2, \ldots, \gamma_p)$ follows a Potts distribution, 
\begin{equation}
p(\boldsymbol{\gamma}|\boldsymbol{\theta}_0, \boldsymbol{\theta}_1) = c(\boldsymbol{\theta}_0, \boldsymbol{\theta}_1)^{-1} \textnormal{exp} \Big\{ \sum_{i=1}^{p} \theta_{0k}I[ \gamma_i = k ] \Big\} \times \textnormal{exp} \Big\{ \sum_{i=1}^{p} \sum_{i \sim j} \sum_{k=1}^{4} \theta_{1k}I[ \gamma_i = \gamma_j = k ] \Big\}
\label{eq:pottspdf}
\end{equation}
where $i \sim j$ indicates neighboring pairs and $I(\cdot)$ is the indicator function. The neighboring relationship can be defined in terms of domain knowledge, or, in our case, the mediator correlation information. $\boldsymbol{\theta}_0 = (\theta_{01}, \theta_{02}, \theta_{03}, \theta_{04})$ effectively determines the four group proportions \textit{a priori} in the absence of mediator correlation. $\boldsymbol{\theta}_1 = (\theta_{11}, \theta_{12}, \theta_{13}, \theta_{14})$ represents how mediator correlation determines the extent to which one mediator being selected into one group affects the probability of its neighboring mediators being selected into the same group. For $\theta_{1k} > 0$, the Potts distribution encourages configurations where ``neighboring mediators" belong to the same group; and the larger $\theta_{1k}$, the tighter this coupling. When $\boldsymbol{\theta}_1 = \boldsymbol{0}$, group membership of one mediator is independent of that of its neighbors. Based on the full probability distribution in Equation \ref{eq:pottspdf}, the probability for the $j$-th mediator belonging to component $k$ conditional on its neighbors is,
\begin{equation}
p(\gamma_j = k | \{ \gamma_i \}_{i \neq j}, \boldsymbol{\theta}_0, \boldsymbol{\theta}_1) = \frac{\textnormal{exp}\{ \theta_{0k}\} \times \textnormal{exp}\{ \sum_{i \sim j} \theta_{1k} I[ \gamma_i = \gamma_j = k ] \}}{\sum_{k = 1}^{4} \textnormal{exp} \{ \theta_{0k}\} \times \textnormal{exp} \{ \sum_{i \sim j} \theta_{1k} I[ \gamma_i = \gamma_j = k ] \} }
\label{eq:conditionalpdf}
\end{equation}
This conditional probability depends on the neighbors of the $j$-th mediator and demonstrates the Markov property of the Potts distribution. 

We develop a Markov chain Monte Carlo (MCMC) sampling strategy for the proposed model. A key challenge for inference is the exact calculation of the normalizing constant $c(\boldsymbol{\theta}_0, \boldsymbol{\theta}_1)$ in Potts distribution, as it requires the summation over the entire space of $\boldsymbol{\gamma}$ which consists of $4^p$ states. Even for a moderate number of mediators, $c(\boldsymbol{\theta}_0, \boldsymbol{\theta}_1)$ is computationally intractable, and this complicates the Bayesian inference. Due to the intractable normalizing constant in Potts distribution, the update of $\boldsymbol{\theta}_0, \boldsymbol{\theta}_1$ cannot be handled by the standard Metropolis Hastings (MH) algorithm. To address this issue, we employ the double MH sampler \citep{liang2010double} to generate auxiliary variables via the MH transition kernels and eliminate the normalizing constants. For $\boldsymbol{\theta}_0, \boldsymbol{\theta}_1$, we consider normal priors, and the prior means of $\{ \theta_{0k} \} $ are set to have the desired inclusion probability while the prior means of $\{ \theta_{1k} \} $ are set to be the same positive number.  This prior information favors the grouping of correlated mediators. According to Equation \ref{eq:conditionalpdf}, the updating of $\boldsymbol{\gamma}$ can be realized through single site Gibbs sampling. Since the sampling space of $\boldsymbol{\gamma}$ is huge and discrete, the efficiency of the standard Gibbs updates can be improved by the Swendsen-Wang (SW) algorithm \citep{higdon1998auxiliary}. The SW algorithm partitions the whole set of mediators into blocks within which the mediators belong to the same normal component, and then updates each block independently. Following the strategy in \cite{higdon1998auxiliary}, we alternate between the single site Gibbs updates of $\boldsymbol{\gamma}$ and SW updates to ensure movement in large patches and fast mixing of the algorithm. The detailed algorithm is given in the SM.

In our Potts mixture model, the ``neighboring" mediators are predefined to capture the correlation structure among mediators. Based on our experience, including too many neighbors into the model will cause irrelevant noises to the group probabilities and blur the cluster boundary; while including too few neighbors will certainly lose some of the important structural information. In this paper, we apply the common clustering method on the $p(p-1)/2$ pairwise correlations across the $p$ mediators to divide them into two groups: high correlation and background noise. This procedure essentially sets a correlation threshold for neighbors and non-neighbors in a data dependent way. In the procedure, we define the $i$-th mediator and $j$-th mediator as neighbors if their pairwise correlation is above this threshold. The threshold may be determined in other ways to reflect the prior knowledge on the neighborhood structure and relationships across mediators. 

We refer to our Potts mixture model as GMM-Potts. GMM-Potts translates the correlation structure into a neighboring graph and incorporates the local dependency among mediators through mediators' predefined neighbors. For each mediator, its four-component group probabilities will be dependent on its neighboring correlated mediators but not the non-neighboring ones. This local dependency feature of GMM-Potts is unique as compared to the previous GMM and does not incur much additional computational burden. 

\subsection{Hierarchical GMM with Correlated Selection: GMM-CorrS}
\label{sec32}
GMM-Potts requires a hard thresholding rule to determine the neighboring graph among mediators. If the neighbors and non-neighbors of mediators are not correctly specified or difficult to specify as in the case of a weak correlation structure, then GMM-Potts may incur a loss of performance. To avoid the need of neighborhood pre-specification and allow for a more direct incorporation of correlation structure, we consider an alternative approach for coordinated selection of correlated mediators here. This alternative approach is again built upon the GMM framework. Specficially, for each mediator, we assume that the selection/group indicator $\gamma_j$ follows a multinomial distribution with parameters $\pi_{1j}$, $\pi_{2j}$, $\pi_{3j}$, $\pi_{4j}$, and $\sum_{k=1}^4\pi_{kj} = 1$. We propose to jointly model all the mediators' mixing probabilities and their continuous dependence structure via latent logistic normal distributions. The logistic normal \citep{atchison1980logistic} has been studied in the context of analyzing compositional data, such as bacterial composition in human microbiome data \citep{xia2013logistic} and topics proportions associated with document collections in correlated topics model \citep{chen2013scalable}. In mediation analysis, it would allow for a flexible covariance structure among mediators and give a more realistic model where correlated mediators will have similar group probabilities \textit{a priori}.

In particular, we employ a P\'olya-Gamma (PG) latent variable representation of the multinomial distribution to enable coordiated mediator selection. Our approach is motivated in part by computational considerations. Specifically, a naive incorporation of the Gaussian correlation structure among multinomial parameters as described in the previous paragraph imposes substantial computational challenge, as it would break the Dirichlet-multinomial conjugacy commonly used in mixture models. Approximation techniques, such as variational inference are feasible, but they do not always come with the theoretical guarantees as MCMC \citep{blei2007correlated}. Our approach extends a similar approach in Bayesian logistic regression inference. Specifically, Bayesian logistic regression has long been explored given its inconvenient analytic form of the likelihood and the non-existence of a conjugate prior for parameters of interest. Recently, \cite{polson2013bayesian} constructs a new data-augmentation strategy based on the novel class of P\'olya-Gamma (PG) distributions, and the method is notably simpler and more efficient than the previous schemes for Bayesian hierarchical models with binomial likelihoods \citep{holmes2006bayesian}. To extend that approach to multinomial logit models and facilitate MCMC computation, we leverage a logistic stick-breaking representation in the PG latent variable augmentation \citep{linderman2015dependent} to formulate the multinomial distribution in terms of latent variables with the jointly Gaussian likelihoods. First, we rewrite 4-dimensional multinomial in terms of 3 binomial densities $\widetilde{\pi}_{j1}$, $\widetilde{\pi}_{j2}$ and $\widetilde{\pi}_{j3}$,
\begin{eqnarray*}
p(\gamma_j = 1) &=& \widetilde{\pi}_{j1} = \pi_{j1} \\
p(\gamma_j = 2 | \gamma_j \neq 1) &=& \widetilde{\pi}_{j2} = \pi_{j2}/(1-\pi_{j1}) \\
p(\gamma_j = 3 | \gamma_j \neq 1 \textnormal{ or } 2) &=& \widetilde{\pi}_{j3} = \pi_{j3}/(1-\pi_{j1}-\pi_{j2}) \\ 
p(\gamma_j = 4 | \gamma_j \neq 1 \textnormal{ or } 2 \textnormal{ or } 3) &=& \widetilde{\pi}_{j4} = \pi_{j4}/(1-\pi_{j1}-\pi_{j2}-\pi_{j3}) = 1 \\
\textnormal{Multinomial}(\gamma_j | 1, \{ \pi_{j1}, \pi_{j2}, \pi_{j3}, \pi_{j4} \} ) &=& \prod_{k=1}^3 \textnormal{Binomial}( I(\gamma_j = k) |  n_{jk}, \widetilde{\pi}_{jk})
\end{eqnarray*}
where $n_{jk} = 1-\sum_{k^{'} < k}I(\gamma_{j} = k^{'})$, $n_{j1} = 1$. The multinomial distribution is now expressed with three binomial distributions and each $\widetilde{\pi}_{jk}$ describes the faction of the remaining probability for the $k$-th group (details in the SM). To better aid the interpretation of the above stick-breaking representation, we may consider a testing strategy for the indirect effect $\beta_{mj}\alpha_{aj}$ implemented on each mediator. By doing that, we will get the subset of active mediators with $\beta_{mj}\alpha_{aj} \neq 0$, i.e. $\gamma_j = 1$. For the remaining mediators with $\beta_{mj}\alpha_{aj} = 0$, we further consider the following three cases: $p(\gamma_j = 2 | \gamma_j \neq 1)$ is the conditional probability of having non-zero $\beta_{mj}$ effect but zero $\alpha_{aj}$ given that $\beta_{mj}\alpha_{aj} = 0$; $p(\gamma_j = 3 | \gamma_j \neq 1 \textnormal{ or } 2)$ is the conditional probability of having non-zero $\alpha_{aj}$ effect given that $\beta_{mj} = 0$; and the rest of the mediators will surely have $\beta_{mj} = \alpha_{aj} = 0$, i.e. $\gamma_j = 4$. We note that under the sparsity assumption, for most of the mediators, $\widetilde{\pi}_{j2} \approx \pi_{j2}$, $\widetilde{\pi}_{j3} \approx \pi_{j3}$ due to the small values of $\pi_{j1}$ and $\pi_{j2}$.

Then, we define $b_{jk} = \textnormal{logit}(\widetilde\pi_{jk})$ for $k = 1,2,3$ and $j = 1, 2, \ldots, p$. We stack the $3 \times p$ $b_{jk}$'s as one random vector, and assume a multivariate normal prior on it, that is,
\begin{equation*}
\boldsymbol{b} := \{b_{jk}\}_{j = 1, \ldots, p; k = 1, 2,3} 
\label{eq:logit}
\end{equation*}
\begin{equation}
\boldsymbol{b} \sim \textnormal{MVN}(\boldsymbol{a}, \textnormal{diag}\{ \sigma_{d1}^2, \sigma_{d2}^2, \sigma_{d3}^2 \} \otimes \boldsymbol{D})
\end{equation}
where $\otimes$ denotes the Kronecker product. The logistic transformation maps the transformed multinomial parameters to the $3p$-dimensional open real space.  The prior mean $\boldsymbol{a} = \{a_{jk}\}_{j = 1, \ldots, p; k = 1,2,3}$, and it is chosen such that $a_{jk} = a_{j'k}$ for $k = 1,2,3$ and $1 \leq j < j' \leq p$. It reflects our prior belief on the overall group proportions and induces sparsity for the first three groups. The $\boldsymbol{D}$ is a $p$-by-$p$ covariance matrix and will incorporate the mediator-wise correlation/structure dependency to the transformed mixing probabilities. In our setting, we estimate the correlation matrix among mediators from data and replace the negative correlations with their absolute values. We then find the nearest positive definite matrix to the absolute correlation matrix, and use that as the $\boldsymbol{D}$ matrix in model fitting. Since the variation level may be different for $\textnormal{logit}(\widetilde{\pi}_{j1})$, $\textnormal{logit}(\widetilde{\pi}_{j2})$ and $\textnormal{logit}(\widetilde{\pi}_{j3})$, we introduce the group-wise $\sigma_{dk}^2, k = 1, 2, 3$ for a more general covariance pattern. This correlation embedded GMM exploits the whole correlation information from all the mediators and does not require the predefined neighbors as in the GMM-Potts model. 

We refer to the above model as GMM-CorrS. We develop an MCMC algorithm to infer parameters through data augmentation with P\'olya-Gamma variables \citep{polson2013bayesian}. The augmented posterior leads to conditional distributions from which we can easily draw samples and the entire vector $\boldsymbol{b}$ can be sampled as a block in a single Gibbs update. The detailed derivation and algorithm can be found in the SM.

\section{Simulations}
\label{sec4}
We evaluate the performance of the proposed models compared with existing methods under different scenarios through simulations. 

\subsection{Small Sample Scenarios: $n = 100, p = 200$}
\subsubsection{Simulation Design}
Following settings in \cite{song2020mediation}, we adopt the four-component structure to generate the exposure-mediator and mediator-outcome effects, i.e. simulate $\large [ 
\beta_{mj}, \alpha_{aj} \large ]^\top$ from 
\begin{equation*}        
[ 
\beta_{mj}, \alpha_{aj} ]^\top \sim \pi_{1}\mathrm{MVN}(\boldsymbol{0}, 		   \colvec{0.5 & 0.2 \\ 0.2 & 0.5}) + \pi_{2}\mathrm{MVN}(\boldsymbol{0}, \colvec{0.5 & 0 \\ 0 & 0}) + \pi_{3}\mathrm{MVN}(\boldsymbol{0}, \colvec{0 & 0 \\ 0 & 0.5}) + \pi_{4}\boldsymbol{\delta}_0
\end{equation*}
To introduce sparsity, we assume the proportion of active mediators $\pi_{1} = 0.05$, and the other three null components $\pi_{2} = 0.05, \pi_{3} = 0.10, \pi_{4} = 0.80$. We generate a $p$-vector of correlated mediators for the $i$th individual from $\boldsymbol{M}_i = A_i\boldsymbol{\alpha_a} + \boldsymbol{\epsilon}_{M_i}$, where the continuous exposure $\{ A_i, i = 1, \ldots, n \} $ is independently sampled from a standard normal distribution. The residual errors $\boldsymbol{\epsilon}_{M_i} \sim \textnormal{MVN}(\boldsymbol{0}, \boldsymbol{\Sigma})$ and ${\boldsymbol{\Sigma}}$ models the correlation structure across mediators. For the outcome, we simulate it from the linear model: $Y_i = \boldsymbol{M}_i^{\top}\boldsymbol{\beta_m} + A_i\beta_a + \epsilon_{Y_i}$, with $\beta_a = 0.5$, and the residual error $\epsilon_{Y_i} \sim \textnormal{N}(0, 1)$.

For the correlation structure, we assume 10 highly-correlated blocks of size 10 $\times$ 10, within which the pairwise correlation of mediators is $\rho_1$, e.g. $\rho_1 = 0.5-0.03|i-j|$ or $0.9-0.05|i-j|$, and the correlation between blocks ($\rho_2$) is relatively weak (e.g. $\rho_2 = 0$ or $0.1$). Such correlation structure mimics the local dependency due to physical adjacency or biologically functional pathway of biomarkers, which is commonly seen in the high-dimensional mediators. There are 10 active mediators, and they are assumed to cluster within one block or scatter over a few blocks, while the other blocks contain no active mediators. We also consider settings where there is no correlation or such structural information underlying active mediators, that is, setting ${\boldsymbol{\Sigma}}$ to be identical matrix or estimated covariance based on a random subset of DNAm from MESA. For the Bayesian methods, we check the MCMC convergence by running ten chains and computing the potential scaled reduction factors (PSRF, \cite{gelman1992inference}). The estimated 95\% confidential interval of the PSRFs for all the PIPs is [1.0, 1.2], indicating good mixing and convergence of the algorithms. 

The GMM-Potts model needs the input of a reliable neighborhood matrix. In practice, we may not be able to specify a completely precise neighborhood structure, but instead a deviated version of that. To examine how sensitive our GMM-Potts model is to the incorrect neighborhood relationship, we randomly convert a proportion of $r$ neighboring mediator pairs to be non-neighboring, and randomly convert the same amount of non-neighboring pairs to be neighbors. The other configurations are the same as in the previous simulations. We vary the perturbation rate $r$ from 0.05 to 0.5 to mimic different degrees of bias. In addition, for the GMM-CorrS, since it directly takes the correlation matrix as an input, we examine its sensitivity to the observed correlation matrix by adding mild changes from $\textnormal{N}(0, \sigma^2)$ to the estimated matrix. We vary $\sigma$ from 0.1 to 0.3 for different levels of noise. 

\subsubsection{Evaluation Metrics}

To examine the mediator selection accuracy, for the proposed GMM-Potts and GMM-CorrS methods as well as GMM, we use PIP to rank and select mediators. We calculate the true positive rate (TPR) for active mediators based on the fixed 10\% false discovery rate (FDR). For the estimation accuracy, we calculate the mean square error (MSE) of the indirect effects for both non-null and null mediators, denoted as MSE$_{\textnormal{non-null}}$ and MSE$_{\textnormal{null}}$. We perform 200 replicates for each scenario and report the means of those metrics in the result tables. 

\subsubsection{Competing Methods} 

In addition to the proposed methods, we consider the following existing methods: GMM with no correlated information included, Bi-Lasso (apply two separate Lasso regressions \citep{tibshirani1996regression} to the outcome and mediator model, respectively), Bi-Ridge (apply two separate ridge regressions \citep{hoerl1988ridge}  to the outcome and mediator model, respectively), and Pathway Lasso \citep{zhao2016pathway}. In Bi-Lasso and Bi-Ridge, we adopt 10-fold cross validation to choose the tuning parameter in each regression separately. The three frequentist methods provide optimized solutions of $\boldsymbol{\beta_m}$, $\boldsymbol{\alpha_a}$ to the three different penalized likelihoods, and the marginal indirect contribution from each mediator, i.e. $\beta_{mj}\alpha_{aj}$, is used to rank mediators for the TPR calculation.

\subsubsection{Simulation Results}

Table \ref{tbl:power1} shows the results under the small sample scenarios with $n=100, p=200$. Overall, by leveraging mediators' correlation structure, the two proposed approaches, GMM-Potts and GMM-CorrS, substantially improve the selection accuracy over the other methods. When the active mediators are concentrated within one block, the GMM-Potts achieves the highest TPR ($>$ 0.90) at a fixed 10\% FDR for identifying this whole block, followed by GMM-CorrS ($\sim$0.80 TPR). The advantage of the proposed methods grows with stronger correlations. Without such ``group selection" ability, the GMM under independent priors tends to lose half of the power for detecting correlated mediators. On the other hand, if the active ones are evenly distributed into two blocks, then highly correlated mediators within the same block may not be concurrently active. This could happen if their correlation does not mainly link with mediation as we assume, and therefore may disturb mediator selection. Under those settings, we do observe power decrease for the proposed methods. Particularly, the GMM-Potts model becomes less preferable as it smoothes over non-mediating neighbors to infer active mediators; while GMM-CorrS uses a more flexible Gaussian distribution for dependent group probabilities and thus has the best TPR. In the settings where there is no systematic correlation structure underlying mediators, we find that GMM-CorrS behaves quite similarly to the GMM, and outperforms the others. GMM-Potts is less robust presumably due to the inclusion of irrelevant neighbors, but still better than the frequentist methods. The three frequentist methods have relatively poor selection performance with highly correlated mediators, and Bi-Lasso is most competitive under zero or weak correlation. In terms of the effects estimation, the proposed methods mostly achieve the smallest MSE$_{\textnormal{non-null}}$ and a reasonable level of MSE$_{\textnormal{null}}$. Among the three frequentist methods, since in general Lasso tends to select less correlated variables than the elastic net type penalty, Bi-Lasso has a relatively larger MSE$_{\textnormal{non-null}}$ but noticeably smaller MSE$_{\textnormal{null}}$ than the pathway Lasso. Given the sparse setup in the above simulations, Bi-Ridge does not exhibit much advantage over the other methods.

\begin{table}[!h]
\scalebox{0.9}{
\centering{
  \begin{tabular}{c  c c c c c c}
    \hline\hline
    & \multicolumn{6}{c}{$\rho_1 = 0.5 - 0.03|i-j|, \rho_2 = 0$} \\
    & \multicolumn{3}{c}{(A) Signals in one block} & \multicolumn{3}{c}{(B) Signals in two blocks} \\
    \hline\hline
    Method & TPR & MSE$_{\textnormal{non-null}}$ & MSE$_{\textnormal{null}}$ $\times 10^{-4}$ & TPR & MSE$_{\textnormal{non-null}}$ & MSE$_{\textnormal{null}}$ $\times 10^{-4}$ \\
    \hline
    GMM-CorrS & \bf{0.78} & 0.029 & 1.360 & \bf{0.62} & 0.039 & 1.919 \\
    GMM-Potts & \bf{0.93} & 0.035 & 2.251 & \bf{0.49} & 0.040 & 2.112 \\
    GMM & 0.45 & 0.042 & 1.211 & 0.46 & 0.047 & 1.203 \\
    Bi-Lasso & 0.26 & 0.238 & 0.520 & 0.23 & 0.238 & 0.584 \\
    Bi-Ridge & 0.22 & 0.283 & 2.639 & 0.21 & 0.286 & 2.642 \\
    Pathway Lasso & 0.24 & 0.233 & 2.598 & 0.23 & 0.180 & 6.405 \\
    \hline\hline
    & \multicolumn{6}{c}{$\rho_1 = 0.9 - 0.05|i-j|, \rho_2 = 0.1$} \\
    & \multicolumn{3}{c}{(A) Signals in one block} & \multicolumn{3}{c}{(B) Signals in two blocks} \\
    \hline\hline
    Method & TPR & MSE$_{\textnormal{non-null}}$ & MSE$_{\textnormal{null}}$ $\times 10^{-4}$ & TPR & MSE$_{\textnormal{non-null}}$ & MSE$_{\textnormal{null}}$ $\times 10^{-4}$ \\
    \hline
    GMM-CorrS & \bf{0.81} & 0.208 & 1.146 & \bf{0.49} & 0.182 & 4.080 \\
    GMM-Potts & \bf{0.92} & 0.171 & 3.515 & \bf{0.41} & 0.233 & 1.651 \\
    GMM  & 0.33 & 0.206 & 2.158 & 0.22 & 0.201 & 3.112 \\
    Bi-Lasso & 0.11 & 0.342 & 0.173 & 0.13 & 0.343 & 0.179 \\
    Bi-Ridge & 0.15 & 0.322 & 2.170 & 0.16 & 0.326 & 1.690\\
    Pathway Lasso & 0.21 & 0.237 & 5.495 & 0.19 & 0.264 & 3.457 \\    
     \hline\hline
    & \multicolumn{6}{c}{No systematic correlation structure (signals in two blocks)} \\
    & \multicolumn{3}{c}{(A) $\rho_1 = 0$} & \multicolumn{3}{c}{(B) Weak correlation from MESA} \\
    \hline\hline
    Method & TPR & MSE$_{\textnormal{non-null}}$ & MSE$_{\textnormal{null}}$ $\times 10^{-4}$ & TPR & MSE$_{\textnormal{non-null}}$ & MSE$_{\textnormal{null}}$ $\times 10^{-4}$ \\
    \hline
    GMM-CorrS & \bf{0.52} & 0.020 & 1.042 & \bf{0.44} & 0.023 & 1.780 \\
    GMM-Potts & 0.46 & 0.043 & 1.970 & 0.40 & 0.030 & 3.041 \\
    GMM & \bf{0.52} & 0.021 & 0.805 & \bf{0.45} & 0.023 & 1.642 \\
    Bi-Lasso & 0.45 & 0.081 & 0.542 & 0.35 & 0.139 & 0.740 \\
    Bi-Ridge & 0.35 & 0.238 & 3.645 & 0.28 & 0.247 & 4.003 \\
    Pathway Lasso & 0.35 & 0.164 & 0.314 & 0.32 & 0.177 & 0.400 \\
    \hline
\end{tabular}}}
\caption{Simulation results of $n=100, p = 200$ under different correlation structures.  TPR:  true positive rate at false discovery rate (FDR) = 0.10. MSE$_{\textnormal{non-null}}$: mean squared error for the indirect effects of active mediators. MSE$_{\textnormal{null}}$: mean squared error for the indirect effects of inactive mediators. The results are based on 200 replicates for each setting. Bolded TPRs indicate the top two performers.}
\label{tbl:power1}
\end{table}

Tables \ref{tbl:sens1} and \ref{tbl:sens2} summarize the sensitivity analysis for GMM-Potts and GMM-CorrS, respectively, regarding the input correlation structure. As expected, with increasing noise added to the correlation structure, the overall accuracy of GMM-Potts and GMM-CorrS gets reduced. However, the power of our methods remains 75\% of the original level for reasonable $r$ and $\sigma$ ($r < 0.3$, $\sigma < 0.3$). Even with large $r = 0.5$ and $\sigma = 0.3$, GMM-CorrS still has better performance (TPR, MSE$_{\textnormal{non-null}}$) over methods with no structural information in all the settings, and GMM-Potts does for most of the settings. Generally speaking, the proposed methods are not sensitive to small alteration of the input correlation structure.

\begin{table}[!h]
\scalebox{0.87}{
\centering{
  \begin{tabular}{c  c c c c c c}
    \hline\hline
    & \multicolumn{6}{c}{$\rho_1 = 0.5 - 0.03|i-j|, \rho_2 = 0$} \\
    & \multicolumn{3}{c}{(A) Signals in one block} & \multicolumn{3}{c}{(B) Signals in two blocks} \\
    \hline\hline
    Perturbation rate & TPR & MSE$_{\textnormal{non-null}}$ & MSE$_{\textnormal{null}}$ $\times 10^{-4}$ & TPR & MSE$_{\textnormal{non-null}}$ & MSE$_{\textnormal{null}}$ $\times 10^{-4}$ \\
    \hline
    0  & 0.93 & 0.035 & 2.251 & 0.49 & 0.040 & 2.112 \\
    0.05 & 0.78 & 0.076 & 1.496 & 0.44 & 0.091 & 1.733 \\
    0.1 & 0.72 & 0.077 & 1.578 & 0.43 & 0.091 & 1.827 \\
    0.2 & 0.69 & 0.087 & 1.568 & 0.42 & 0.086 & 1.822 \\
    0.3 & 0.61 & 0.097 & 1.736 & 0.41 & 0.088 & 2.019 \\
    0.4 & 0.53 & 0.102 & 1.525 & 0.40 & 0.085 & 1.952 \\
    0.5 & 0.49 & 0.094 & 2.082 & 0.41 & 0.081 & 1.847 \\
    \hline\hline
    & \multicolumn{6}{c}{$\rho_1 = 0.9 - 0.05|i-j|, \rho_2 = 0.1$} \\
    & \multicolumn{3}{c}{(A) Signals in one block} & \multicolumn{3}{c}{(B) Signals in two blocks} \\
    \hline\hline
    Perturbation rate & TPR & MSE$_{\textnormal{non-null}}$ & MSE$_{\textnormal{null}}$ $\times 10^{-4}$ & TPR & MSE$_{\textnormal{non-null}}$ & MSE$_{\textnormal{null}}$ $\times 10^{-4}$ \\
    \hline
    0  & 0.92 & 0.171 & 3.515 & 0.41 & 0.233 & 1.651 \\
    0.05 & 0.91 & 0.180 & 0.819 & 0.33 & 0.191 & 1.876 \\
    0.1 & 0.91 & 0.181 & 1.203 & 0.35 & 0.183 & 2.156 \\
    0.2 & 0.91 & 0.175 & 1.393 & 0.32 & 0.201 & 1.815 \\
    0.3 & 0.89 & 0.174 & 1.129 & 0.32 & 0.177 & 2.081 \\
    0.4 & 0.88 & 0.173 & 1.395 & 0.32 & 0.200 & 1.492 \\
    0.5 & 0.83 & 0.166 & 2.046 & 0.30 & 0.188 & 1.884 \\
    \hline
\end{tabular}}}
\caption{Sensitivity analysis for Potts mixture model (GMM-Potts) for $n=100, p = 200$.}
\label{tbl:sens1}
\end{table}

\begin{table}
\scalebox{0.9}{
\centering{
   \begin{tabular}{c  c c c c c c}
    \hline\hline
    & \multicolumn{6}{c}{$\rho_1 = 0.5 - 0.03|i-j|, \rho_2 = 0$} \\
    & \multicolumn{3}{c}{(A) Signals in one block} & \multicolumn{3}{c}{(B) Signals in two blocks} \\
    \hline\hline
    Noise level & TPR & MSE$_{\textnormal{non-null}}$ & MSE$_{\textnormal{null}}$ $\times 10^{-4}$ & TPR & MSE$_{\textnormal{non-null}}$ & MSE$_{\textnormal{null}}$ \\
    \hline
    0  & 0.78 & 0.029 & 1.360 & 0.62 & 0.039 & 1.919 \\
    0.1 & 0.71 & 0.029 & 2.481 & 0.56 & 0.036 & 2.246 \\
    0.2 & 0.60 & 0.031 & 2.575 & 0.50 & 0.037 & 2.043 \\
    0.3 & 0.53 & 0.033 & 2.235 & 0.47 & 0.037 & 1.910 \\
    \hline\hline
    & \multicolumn{6}{c}{$\rho_1 = 0.9 - 0.05|i-j|, \rho_2 = 0.1$} \\
    & \multicolumn{3}{c}{(A) Signals in one block} & \multicolumn{3}{c}{(B) Signals in two blocks} \\
    \hline\hline
    Noise level & TPR & MSE$_{\textnormal{non-null}}$ & MSE$_{\textnormal{null}}$ $\times 10^{-4}$ & TPR & MSE$_{\textnormal{non-null}}$ & MSE$_{\textnormal{null}}$ $\times 10^{-4}$ \\
    \hline
    0  & 0.81 & 0.208 & 1.146 & 0.49 & 0.182 & 4.080 \\
    0.1 & 0.72 & 0.168 & 4.017 & 0.40 & 0.127 & 3.288 \\
    0.2 & 0.63 & 0.170 & 3.442 & 0.37 & 0.130 & 3.370 \\
    0.3 & 0.54 & 0.176 & 3.413 & 0.34 & 0.133 & 3.283 \\
    \hline
\end{tabular}}}
\caption{Sensitivity analysis for the Gaussian mixture model with correlated selection (GMM-CorrS) for $n=100, p = 200$. }
\label{tbl:sens2}
\end{table}

\subsection{Large Sample Scenarios: $n = 1000, p = 2000$}
\subsubsection{Simulation Design}

Next, we examine the settings for $n = 1000, p=2000$. We simulate the exposure, exposure-mediator and mediator-outcome effects using the same distribution as above. For the correlation structure, we now consider 50 blocks of size 20 $\times$ 20, with relatively high within-block mediator correlation $\rho_1$ and zero between-block correlation. We first set the four group proportions same as in the small sample scenarios, and the resultant 100 active mediators are assumed to evenly distribute over five blocks. The other blocks contain no active mediators. In one of the settings, we use the  covariance matrix estimated from a random subset of DNAm in MESA as ${\boldsymbol{\Sigma}}$ to simulate mediators with no underlying systematic correlation structure.

Then we study a much sparser setting with only 10 active mediators to better reflect the situation we observe in the MESA application. The 10 active mediators exist in two blocks, each of which contains five active ones and 15 inactive ones. Furthermore, we consider another worse-case scenario for GMM-Potts model by reducing $\rho_1 $ to 0.25 and remaining the high sparsity. The weak correlation makes it hard for GMM-Potts model to identify the true neighboring relationship via the clustering method, and the performance of the Potts model is quite dependent on the smoothing effects from the predefined neighbors.

\subsubsection{Simulation Results}

Table \ref{tbl:power2} shows the results under the large sample scenarios with $n=1000, p=2000$. Our methods enjoy up to 30\% power gain on mediator selection utilizing the correlation structure compared to the other methods. In the first setting, both methods identify almost all the active blocks, and GMM-Potts has a slightly higher TPR (0.97) at 10\% FDR than GMM-CorrS (TPR = 0.92). When the mediator correlation has no implication for mediation effects in the second setting, the overall performance of GMM-CorrS is similar to that of GMM, and better than GMM-Potts. Those patterns are consistent with what we have observed in the small sample scenarios. Under the much sparser settings with only 10 active mediators and varied correlation $\rho_1 $, the GMM-CorrS maintains good and stable performance with TPR around 0.80. By contrast, the performance of GMM-Potts is dependent on how obvious the correlation patterns are and subsequently how well the clustering method does in defining neighbors and non-neighbors. For example, with $\rho_1 = 0.5 - 0.02|i-j|$, the GMM-Potts models can accurately identify the underlying correlation structure and achieve the highest TPR (0.85), smallest MSE (MSE$_{\textnormal{non-null}}$ = 0.002, MSE$_{\textnormal{null}}$ = 7.607 $\times 10^{-7}$). However, as the within-block correlation $\rho_1$ reduces to 0.25, it becomes challenging for the clustering method to separate true correlation versus noise, and we do observe many noisy pairs in the neighborhood matrix. As a consequence, the results of GMM-Potts model get compromised by the inclusion of those irrelevant neighbors. This setting is actually in agreement with our observation of the ambiguous correlation structure and sparse signals in the MESA application, which may not fare well for GMM-Potts model. Among the other three frequentist methods, Bi-Lasso performs best regarding to the selection and estimation accuracy. 

\begin{table}[!h]
\scalebox{0.9}{
\centering{
  \begin{tabular}{c  c c c c c c}
    \hline\hline
    & \multicolumn{6}{c}{$p_{11} = 100$, Signals in five blocks} \\
    & \multicolumn{3}{c}{(A) $\rho_1 = 0.5 - 0.02|i-j|$} & \multicolumn{3}{c}{(B) Weak correlation from MESA} \\
    \hline\hline
    Method & TPR & MSE$_{\textnormal{non-null}}$ & MSE$_{\textnormal{null}}$ $\times 10^{-4}$ & TPR & MSE$_{\textnormal{non-null}}$ & MSE$_{\textnormal{null}}$ $\times 10^{-4}$ \\
    \hline
    GMM-CorrS & \bf{0.92} & 0.031 & 0.440 & \bf{0.83} & 0.002 & 0.240 \\
    GMM-Potts & \bf{0.97} & 0.030 & 0.018 & 0.76 & 0.004 & 1.013 \\
    GMM & 0.76 & 0.077 & 0.630 & \bf{0.84} & 0.002 & 0.176 \\
    Bi-Lasso & 0.73 & 0.031 & 0.199 & 0.65 & 0.042 & 0.446 \\
    Bi-Ridge & 0.32 & 0.244 & 2.680 & 0.36 & 0.202 & 3.795 \\
    Pathway Lasso & 0.44 & 0.112 & 1.162 & 0.42 & 0.107 & 1.427 \\
    \hline\hline
    & \multicolumn{6}{c}{$p_{11} = 10$, Signals in two blocks} \\
    & \multicolumn{3}{c}{(A) $\rho_1 = 0.5 - 0.02|i-j|$} & \multicolumn{3}{c}{(B) $\rho_1 = 0.25$} \\
    \hline\hline
    Method & TPR & MSE$_{\textnormal{non-null}}$ & MSE$_{\textnormal{null}}$ $\times 10^{-4}$ & TPR & MSE$_{\textnormal{non-null}}$ & MSE$_{\textnormal{null}}$ $\times 10^{-4}$ \\
    \hline
    GMM-CorrS & \bf{0.83} & 0.003 & 0.015 & \bf{0.82} & 0.002 & 0.017 \\
    GMM-Potts & \bf{0.85} & 0.002 & 0.008 & 0.61 & 0.018 & 0.228 \\
    GMM & 0.80 & 0.003 & 0.013 & \bf{0.81} & 0.002 & 0.016 \\
    Bi-Lasso & 0.73 & 0.013 & 0.036 & 0.76 & 0.010 & 0.035 \\
    Bi-Ridge & 0.41 & 0.061 & 1.508 & 0.39 & 0.063 & 1.517 \\
    Pathway Lasso & 0.55 & 0.046 & 0.133 & 0.56 & 0.047 & 0.141 \\
    \hline
\end{tabular}}}
\caption{Simulation results of $n=1000, p = 2000$ under different correlation structures, $p_{11}$ is the number of true active mediators. TPR:  true positive rate at false discovery rate (FDR) = 0.10. MSE$_{\textnormal{non-null}}$: mean squared error for the indirect effects of active mediators. MSE$_{\textnormal{null}}$: mean squared error for the indirect effects of inactive mediators. The results are based on 200 replicates for each setting. Bolded TPRs indicate the top two performers.}
\label{tbl:power2}
\end{table}

We note that the TPR results shown in the above tables represent the best selection performances one can achieve with the proposed methods, as we know the underlying true signals and can perfectly specify the 10\% FDR thresholds. But that is not the case with real data applications.  Therefore, we examine the empirical FDR estimates using (a) the local FDR approach \citep{efron2007size} for a targeted 10\% FDR, (b) median PIP cutoff, and (c) 0.90 PIP cutoff, along with the corresponding TPR estimates. Detailed procedure and the empirical estimates are provided in the SM. Under the small sample scenarios (Table S1), the local FDR approach provides decent and well-controlled empirical FDR for both of the proposed methods, while the estimates by median PIP cutoff and 0.90 PIP cutoff tend to be either slightly overestimated or very conservative. Under the large sample scenarios (Table S2), the local FDR approach and median PIP cutoff still produces reasonable FDR estimates for GMM-CorrS across different settings and for GMM-Potts when neighbors reflect connected signals. However, including irrelevant neighbors in GMM-Potts could lead to increased false discoveries, and instead a more stringent 0.90 PIP cutoff may be used if one seeks a lower limit on the false discovery. Therefore in practice, we would recommend the local FDR and 0.90 PIP cutoff for reasonable FDR estimates and control, and we recognize the potential caveat concerning inflated FDR for GMM-Potts. 

To summarize our findings from the simulations, GMM-CorrS takes the overall correlation structure among mediators directly into the modeling process, and shows excellent performance and robustness under different correlation structures. On the other hand, the performance of GMM-Potts is related to how well the pre-specified neighborhood matrix reflects the underlying connection of active mediators. When the correlation-based neighboring relationship has good implication on similar mediation effects, GMM-Potts usually achieves the best selection and estimation accuracy. Its performance will likely get compromised by the inclusion of irrelevant neighbors.

\section{Data Application}
\label{sec5}
In this section, we study two real data applications of the proposed methods: the LIFECODES birth cohort and the MESA cohort. These two data sets have different correlation strength among mediators and thus can serve to demonstrate the advantages of each of the proposed methods. Specifically, in the LIFECODES birth cohort, the biomarkers present a relatively clear correlation/neighborhood structure. We thus expect GMM-Potts model to work well based on our observation from simulations. On the other hand, the correlation structure in the MESA cohort is relatively weak. We thus expect a better performance from GMM-CorrS as compared to GMM-Potts there. 

\subsection{The LIFECODES Birth Cohort}
\begin{figure}[!h]
\begin{center}
\includegraphics[scale=0.45]{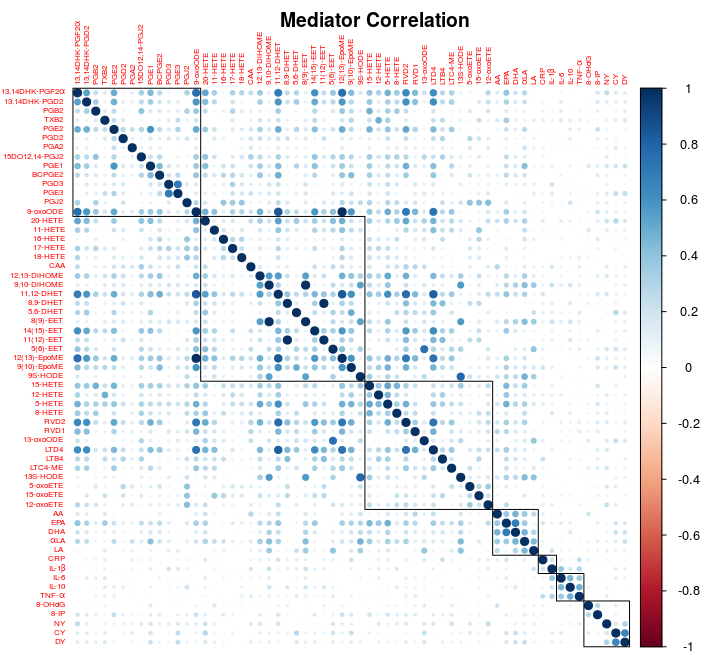}
\end{center}
\caption[]{Correlations among biomarkers in LIFECODES birth cohort. The negative correlations ($\sim$37\% of all the pairwise correlations) were replaced with their absolute values. The 61 biomarkers were grouped by literature derived  biological pathways or processes (black lines).}
\label{fig:correlation}
\end{figure}

In this application, we consider a set of $n = 161$ pregnant women registered at the Brigham and Women's Hospital in Boston, MA between 2006 and 2008. We are interested in the mediation mechanism linking environmental contaminant exposure during pregnancy to preterm birth through endogenous signaling molecules. Those endogenous biomarkers are derived from lipids, peptides, and DNA, and the lipids and peptide derived biomarkers were measured from subjects' plasma samples, while the oxidative stress markers of DNA damage were measured from subjects' urine samples. Both the urine and plasma specimens were collected at one study visit between 23.1 and 28.9 weeks gestation. We focus on $p = 61$ available endogenous biomarkers as potential mediators, including 51 eicosanoids, five oxidative stress biomarkers and five immunological biomarkers. The correlation structure across mediators are shown in Figure \ref{fig:correlation}, and clear pattern with moderate to strong correlations can be observed. For the prenatal exposure to environmental toxicants, we focus the attention of this present study on one class of environmental contaminants, polycyclic aromatic hydrocarbons (PAHs).  PAHs are a group of organic contaminants that form due to the incomplete combustion of hydrocarbons, and commonly present in tobacco smoke, smoked and grilled food products, polluted water and soil, and vehicle exhaust gas \citep{alegbeleye2017polycyclic}. Previous studies have suggested association between PAH exposure and adverse birth outcomes \citep{padula2014exposure}. Since the PAH class contains multiple chemical analytes in our study, we follow \cite{Aung2020.05.30.20117655} to construct an environmental risk score for the PAH class and use that risk score as the exposure variable. The continuous birth outcome, gestational age, was recorded at delivery for each participant, and preterm is defined as delivery prior to 37 weeks gestation. Since the cohort is oversampled for preterm cases, we multiply the data by the case-control sampling weights to adjust for that. We log-transform all measurements of the exposure metabolites and endogenous biomarkers.  We apply the proposed methods with the aforementioned exposure, mediator and outcome variables, controlling for age and maternal BMI from the initial visit, race, and urinary specific gravity levels in both regressions of the mediation analysis.

The results are summarized in Table \ref{tbl:data1}. Based on 10\% FDR using the local FDR approach, GMM-Potts identifies four biomarkers for actively mediating the impact of PAH exposure on gestational age at delivery, 8,9-epoxy-eicosatrienoic acid (8(9)-EET), 9,10-dihydroxy-octadecenoic acid (9,10-DiHOME), 12,13-epoxy-octadecenoic acid (12(13)-EpoME), 9-oxooctadeca-dienoic acid (9-oxoODE); while both GMM-CorrS and GMM only identifies two of them, 8(9)-EET and 9,10-DiHOME. We also report the indirect effect estimates and their 95\% credible intervals for selected mediators, and the direction of effects are consistent among different methods. Among the four biomarkers, 8(9)-EET, 9,10-DiHOME and 12(13)-EpoME belong to the same Cytochrome p450 (CYP450) Pathway; while 9-oxoODE is within Cyclooxygenase (COX) Pathway. CYP450 is a family of enzymes that function to metabolize environmental toxicants, drugs, and endogenous compounds \citep{sadler2016hepatic}, and thus the PAH exposure may cause perturbations in the functions of these enzymes. It has also been suggested that the group of CYP450 metabolites as well as the related genes may play a role in the etiology of preterm delivery \citep{banerjee2014assessment}, and the underlying mechanisms involve increased maternal oxidative stress and inflammation \citep{ferguson2017environmental}. This evidence helps explain the potential mediating mechanism of CYP450 metabolites from PAH exposure to preterm delivery. Additionally, single biomarker analysis also demonstrated the protective effect of 12(13)-EpoME on preterm \citep{aung2019prediction}. We also performed the posterior predictive checks on the outcome model for the three methods, in which the data generated from the posterior predictive distribution are compared with the observed outcome. We find the Bayesian predictive $P$-values \citep{neelon2010bayesian} of the GMM-Potts model are 0.72 and 0.48 for sample first and second moments, respectively, which are closest to 0.5 among the three methods and indicate the most adequate fit of the outcome model. 

Besides the estimated correlation structure, we also consider the input of biological pathway based structural information. That is, only mediators within the same literature derived biological pathway or process are treated as neighbors in GMM-Potts and have non-zero pairwise correlations in GMM-CorrS. The findings are shown in Table S6 of the SM. GMM-Potts identifies a subset of the above four biomarkers: 8(9)-EET, 9,10-DiHOME, and GMM-CorrS declares the other two biomarkers as active mediators: 12(13)-EpoME, 9-oxoODE. The overlapping lists of active mediators add confidence to our findings, and also reveal the fact that only adjusting for biological pathways may lose the correlated information between different pathways.

\begin{table}[!h]
\centering{
   \begin{tabular}{ cccc }
   \hline
	Method & Selected Mediators & PIP & $\hat{\beta}_{mj}\hat{\alpha}_{aj}$ (95\% CI)\\
    \hline
    \multicolumn{4}{c}{\textit{ Polycyclic aromatic hydrocarbons $\rightarrow$ Biomarkers $\rightarrow$ \textit{Gestational Age}}} \\
    \hline
     GMM-Potts & \textcolor{blue}{12(13)-EpoME} & \textcolor{blue}{0.99} & \textcolor{blue}{0.419(0.295, 0.579)} \\
     & 8(9)-EET & 0.98 & 0.368(0.179, 0.567) \\
     & \textcolor{blue}{9-oxoODE} & \textcolor{blue}{0.97} & \textcolor{blue}{-0.296(-0.441, 0.000)} \\
     & 9,10-DiHOME & 0.87 & -0.185(-0.383, 0.000) \\
    \hline
\end{tabular}
}
\caption{Summary of the identified active mediators from the data application on LIFECODES study based on 10\% FDR with the local FDR approach. Compared to GMM-CorrS and GMM, the GMM-Potts model achieves the most adequate fit of the outcome model based on posterior predictive check. The two additional findings from GMM-Potts are marked in blue. Besides the PIP, we also report the effect estimation $\hat{\beta}_{mj}\hat{\alpha}_{aj}$ and its 95\% credible interval (CI). }
\label{tbl:data1}
\end{table}

\subsection{The MESA Cohort}
In this application, we study the mediation mechanism of DNAm in the pathway from neighborhood socioeconomic disadvantage to blood glucose. We focus on $n=1226$ participants with no missing data, and a subset of $p=2000$ CpG sites that have the strongest marginal associations with neighborhood disadvantage for computational reasons. As the exposure, neighborhood socioeconomic disadvantage evaluates the neighborhood social conditions from dimensions of education, occupation, income and wealth, poverty, employment, and housing. Previous literature has demonstrated the relationship between DNA methylation patterns and socially patterned stressors including low adult socioeconomic status (SES) \citep{needham2015life} and unfavorable neighborhood conditions \citep{smith2017neighborhood}. It has also been long known that disadvantaged neighborhood conditions can lead to a variety of health problems, such as chronic psychological distress \citep{ross2009neighborhood} and increased risk of cardiovascular disease \citep{kaplan1993socioeconomic}. The outcome, glucose, is one of the most important blood parameters and should be kept within a safe range in order to support vital body functions and reduce the risk of diabetes and heart disease \citep{sasso2004glucose}. Multiple evidence has supported the association between differential DNAm patterns and glucose metabolism \citep{kriebel2016association}. However, the underlying molecular mechanisms that link neighborhood conditions to physical health profiles are not fully elucidated. To take a step forward, we apply the proposed methods for high-dimensional mediation analysis on DNAm. In the outcome model, we adjust for age, gender, race/ethnicity, childhood SES and adult SES (more details on the SES variables can be found at \cite{smith2017neighborhood}). In the mediator model, we control for age, gender, race/ethnicity, childhood SES, adult SES, and enrichment scores for 4 major blood cell types (neutrophils, B cells, T cells and natural killer cells) to account for potential contamination by non-monocyte cell types. All the continuous variables are standardized to have zero mean and unit variance. In general, the correlation among DNAm in our study is relatively weak, and only 3\% of  DNAm pairs have correlation larger than 0.2.

The results can be found in Table \ref{tbl:data2}. Because of the relatively ambiguous correlation structure observed across mediators in MESA, we do not expect big improvement from our methods. Indeed, the GMM-CorrS identifies one more CpG site as active mediators compared to GMM, and three other CpG sites are detected by both GMM-CorrS and GMM. The rank correlation for the mediator rank lists obtained from the two methods is 0.74, indicating the high consistency between them. The indirect effect estimates from the GMM-CorrS are also close to those from the GMM. The one additional finding of CpG site by GMM-CorrS, cg27090988, is close to the gene \textit{OGG1}. This gene, which is involved in the repair of oxidative DNA damage, has been shown up-regulated in type 2 diabetic islet cell mitochondria, and studies have suggested a crucial role of oxidative DNA damage in the pathogenesis of type 2 diabetes (T2D) \citep{tyrberg2002islet, pan2007oxidative}. We also examine the nearby genes to the other three jointly selected CpG sites. Among them, \textit{MYBPC3} is a known cardiomyopathy gene \citep{dhandapany2009common}, and the increased risk of cardiac hypertrophy and heart failure is likely to alter the glucose metabolism \citep{tran2019glucose}; the expression level of \textit{CD101}, a protein involved in innate immunity, was found associated with T2D in a Mendelian randomization analysis \citep{xue2018genome}. As shown in the simulations, GMM-Potts is not quite suitable for a weak correlation structure as in the MESA data, and the method does not identify any active mediators based on 10\% FDR.

\begin{table}[!h]
\centering{
   \begin{tabular}{ ccccc }
   \hline
	Method & Selected Mediators & Nearby Genes & PIP & $\hat{\beta}_{mj}\hat{\alpha}_{aj}$ (95\% CI)\\
    \hline
    \multicolumn{5}{c}{\textit{Neighborhood SES $\rightarrow$ Biomarkers $\rightarrow$ \textit{Glucose}}} \\
    \hline
     GMM-CorrS &  cg19515398 & EIF2C2 & 0.97 & -0.013(-0.026, 0.000) \\
     & cg04000940 & MYBPC3 & 0.96 &  0.016(0.000, 0.029) \\
     & cg17907003 & CD101 & 0.88 & 0.016(0.000, 0.034) \\
      & \textcolor{blue}{cg27090988} & \textcolor{blue}{OGG1} & \textcolor{blue}{0.84} & \textcolor{blue}{-0.011(-0.024, 0.000)} \\
    \hline
\end{tabular}
}
\caption[Summary of the identified active mediators from the data application on MESA study.]{Summary of the identified active mediators from the data application on MESA study based on 10\% FDR using the local FDR approach. We include the nearby gene, PIP, the effect estimation $\hat{\beta}_{mj}\hat{\alpha}_{aj}$ and its 95\% credible interval (CI) for each selected CpG site. The one additional finding from GMM-CorrS is marked in blue. The GMM-Potts does not identify any active mediators based on 10\% FDR. }
\label{tbl:data2}
\end{table}

\section{Discussion}
\label{sec6}
In this paper, we present two hierarchical Bayesian approaches to incorporating the correlation structure across mediators in high-dimensional mediation analysis: (1) through a logistic normal for mixing probabilities (GMM-CorrS), or (2) through a Potts distribution on the group indicators (GMM-Potts). The consequent ``non-separable" priors of both methods inform the grouping and selection of correlated mediators under the composite structure of mediation. The simulation studies show that utilizing the correlation pattern in active mediators, the proposed methods greatly enhance the selection and estimation accuracy over the methods that do not account for such correlation, and maintain decent and comparable performance under no obvious or mis-specified correlation structure. In addition, the analysis on the LIFECODES birth cohort and MESA cohort indicates that our methods can promote the detection of new active mediators, which may have important implications on future research in targeted interventions for preterm birth and diabetes.

Between the two proposed methods, the GMM-Potts tends to perform better when the correlation pattern is obvious and the included neighbors are informative for inference to limit false positives, while the GMM-CorrS enjoys robust performance under various correlation structures. There are several limitations of the proposed methods. First, for GMM-CorrS, it requires the inversion of a $p \times p$ matrix in each iteration of the sampling algorithm, and as $p$ increases to the scale of hundreds of thousands, that step could become the computational bottleneck of the method. Techniques on matrix approximation or fast parallel matrix inversion will be required to speed up the computing time and reduce the memory footprint. Second, for GMM-Potts, smoothing over arbitrary or inaccurately specified neighbors may have a negative effect on its performance, and this can be further improved by imposing adaptive weight for each neighbor to reflect their relative importance. Moreover, the method can be extended to allow for simultaneous inference of both the active mediators and the neighborhood/network structure linking them. In that way, the neighborhood/network structure among mediators does not need to be known \textit{a priori}.

As promising directions for future work, we note that there may be other ways to incorporate mediators' correlation into the modeling process. Recently, testing the multivariate mediation effects from groups of potential mediators has received growing attention \citep{djordjilovic2019global}, and the variance component tests developed by \cite{huang2019variance} can naturally take into account the correlation within groups. Also, \cite{bobb2015bayesian} develops a Bayesian kernel machine regression to incorporate the structure of the multi-pollutant mixtures into the hierarchical model. Those methodologies may provide insightful perspective to applying correlation kernels under the global testing setup in the context of high-dimensional mediation analysis.


\section*{Acknowledgments}

This work was supported by NSF DMS1712933 (B.M., X.Z.), NIH R01HG009124 (X.Z.), NIH R01HL141292 (J.S.), and NIH R01MD011724 (B.N.). MESA and the MESA SHARe project are conducted and supported by the National Heart, Lung, and Blood Institute (NHLBI) in collaboration with MESA investigators. Support for MESA is provided by contracts HHSN268201500003I, N01-HC-95159, N01-HC-95160, N01-HC-95161, N01-HC-95162, N01-HC-95163, N01-HC-95164, N01-HC-95165, N01-HC-95166, N01-HC-95167, N01-HC-95168, N01-HC-95169, UL1-TR-000040, UL1-TR-001079, UL1-TR-001420, UL1-TR-001881, and DK063491. The MESA Epigenomics \& Transcriptomics Study was funded by NHLBI, NIA, and NIDDK grants: 1R01HL101250, R01 AG054474, and R01 DK101921. The authors thank the other investigators, the staff, and the participants of the MESA study for their valuable contributions. A full list of participating MESA investigators and institutions can be found at http://www.mesa-nhlbi.org.

\bibliographystyle{ba}
\bibliography{reference}

\end{document}


\date{}
\maketitle

\section{Identifiability Assumptions for Causal Mediation Analysis}
\hfill \break
We use the same counterfactual notation as in the main manuscript. To connect potential variables to observed data, we make the Stable Unit Treatment Value Assumption (SUTVA) \citep{rubin1980randomization, rubin1986comment}. Specifically, the SUTVA assumes there is no interference between subjects and the consistency assumption, which states that the observed variables are the same as the potential variables corresponding to the actually observed treatment level, i.e., $\boldsymbol{M_i}=\sum_a { \boldsymbol{M_i}(a) I(A_i=a)}$, and $Y_i=\sum_a \sum_{\boldsymbol{m}} {Y_i (a, \boldsymbol{m}) I(A_i = a, \boldsymbol{M_i} = \boldsymbol{m})}$, where $I(\cdot)$ is the indicator function. 

Causal effects are formally defined in terms of potential variables which are not necessarily observed, but the identification of causal effects must be based on observed data. Therefore further assumptions regarding the confounders are required for the identification of causal effects in mediation analysis \citep{vanderweele2014mediation}. We will use $A \indep B | C$ to denote that $A$ is independent of $B$
conditional on $C$. To estimate the average NDE and NIE from observed data, the following assumptions are needed: (1) $Y_i(a,\boldsymbol{m}) \indep A_i | \boldsymbol{C_i}$, no unmeasured confounding for exposure-outcome relationship; (2) $Y_i(a,\boldsymbol{m}) \indep \boldsymbol{M}_i | \{\boldsymbol{C_i}, A_i\}$, no unmeasured confounding for any of mediator-outcome relationship after controlling for the exposure; (3) $\boldsymbol{M}_i(a) \indep A_i | \boldsymbol{C_i}$, no unmeasured confounding for the exposure effect on all the mediators; (4) $Y_i(a,\boldsymbol{m}) \indep \boldsymbol{M}_i(a^{\star}) | \boldsymbol{C_i}$, no downstream effect of the exposure that confounds any mediator-outcome relationship. The four assumptions are required to hold with respect to the whole set of mediators. Finally, as in all mediation analysis, the temporal ordering assumption also needs to be satisfied, i.e., the exposure precedes the mediators, and the mediators precede the outcome.

\section{Posterior Sampling Algorithm Details for GMM-Potts} 
\hfill \break
\textit{Sampling $\begin{bmatrix}
(\boldsymbol{\beta_m})_j \\
(\boldsymbol{\alpha_a})_j \end{bmatrix}$ and $\gamma_{j}$}
\begin{equation*}
    \textnormal{log}p(\begin{bmatrix}
(\boldsymbol{\beta_m})_j \\
(\boldsymbol{\alpha_a})_j \end{bmatrix} | \gamma_{j} = k, .) \propto -\frac{1}{2}\begin{bmatrix}
(\boldsymbol{\beta_m})_j \\
(\boldsymbol{\alpha_a})_j \end{bmatrix}^\top (\boldsymbol{W_j} + \boldsymbol{V_k}^{-1})\begin{bmatrix}
(\boldsymbol{\beta_m})_j \\
(\boldsymbol{\alpha_a})_j \end{bmatrix} + \boldsymbol{w_j}^\top\begin{bmatrix}
(\boldsymbol{\beta_m})_j \\
(\boldsymbol{\alpha_a})_j \end{bmatrix}
\end{equation*}
where $\boldsymbol{W_j} = \textnormal{diag}\{
\sum_{i=1}^n (\sigma_e^2)^{-1} M_{ij}^2, \sum_{i=1}^n (\sigma_g^2)^{-1}A_i^2 \}$, and \\
$\boldsymbol{w_j} = (\sum_{i=1}^n (\sigma_e^2)^{-1}(Y_i - A_i\beta_a - \sum_{j^{'} \neq j}M_{ij^{'}}(\boldsymbol{\beta_m})_{j^{'}} )M_{ij}, \sum_{i=1}^n \boldsymbol{\Sigma}^{-1}M_{ij}A_i )^\top$
\begin{equation*}
    p(\begin{bmatrix}
(\boldsymbol{\beta_m})_j \\
(\boldsymbol{\alpha_a})_j \end{bmatrix} | \gamma_{j} = k, .) \sim \textnormal{MVN}_2((\boldsymbol{W_j} + \boldsymbol{V_k}^{-1})^{-1}\boldsymbol{w_j}, (\boldsymbol{W_j} + \boldsymbol{V_k}^{-1})^{-1}) 
\end{equation*}
\begin{equation*}
    \textnormal{log}p(\gamma_{j} = k|.) \propto -\frac{1}{2}\textnormal{log}|\boldsymbol{W_j}\boldsymbol{V_k} + \boldsymbol{I_2}| + \frac{1}{2}\boldsymbol{w_j}^\top(\boldsymbol{W_j} + \boldsymbol{V_k}^{-1})^{-1}\boldsymbol{w_j} + \theta_{0k} + \sum_{i \sim j} \theta_{1k} I[ \gamma_i = \gamma_j = k ]
\end{equation*}

\textit{Sampling $\boldsymbol{V_k}$}
\begin{eqnarray*}
\textnormal{log}p(\boldsymbol{V_k}|.) &\propto& -\frac{1}{2}(\sum_{j=1}^p I[\gamma_j = k ]+df+d+1)\textnormal{log}|\boldsymbol{V_k}| - \frac{1}{2}tr(\boldsymbol{\Psi_0}\boldsymbol{V_k}^{-1}) \\
&& + \sum_{j=1}^p I[\gamma_j = k ](-\frac{1}{2}\begin{bmatrix}
(\boldsymbol{\beta_m})_j \\
(\boldsymbol{\alpha_a})_j \end{bmatrix}^\top\boldsymbol{V_k}^{-1}\begin{bmatrix}
(\boldsymbol{\beta_m})_j \\
(\boldsymbol{\alpha_a})_j \end{bmatrix}) 
\end{eqnarray*}
\begin{equation*}
    p(\boldsymbol{V_k}|.) \sim \textnormal{Inv-Wishart} (\boldsymbol{\Psi_0} + \sum_{j=1}^p I[\gamma_j = k ] \begin{bmatrix}
(\boldsymbol{\beta_m})_j \\
(\boldsymbol{\alpha_a})_j \end{bmatrix}\begin{bmatrix}
(\boldsymbol{\beta_m})_j \\
(\boldsymbol{\alpha_a})_j \end{bmatrix}^\top, \sum_{j=1}^p I[\gamma_j = k ]+df)
\end{equation*}

\textit{Sampling $\beta_a$}
\begin{equation*}
    \textnormal{log}p(\beta_{a}| .) \propto  -\frac{\beta_{a}^2}{2\sigma_a^2} - \sum_{i=1}^n \{ \frac{(A_i\beta_{a})^2}{2\sigma_e^2} - \sigma_e^{-2}A_{i}(Y_i - \boldsymbol{M_i}^\top\boldsymbol{\beta_m} - \boldsymbol{C_i}^\top\boldsymbol{\beta_c})\beta_{a} \}
\end{equation*}
\begin{equation*}
    p(\beta_{a}| .) \sim N(\frac{\sum_{i=1}^n A_{i}(Y_i - \boldsymbol{M_i}^\top\boldsymbol{\beta_m} - \boldsymbol{C_i}^\top\boldsymbol{\beta_c})}{\sigma_e^2/\sigma_a^2 + \sum_{i=1}^n A_{i}^2}, \frac{1}{1/\sigma_a^2 + \sum_{i=1}^n A_{i}^2/\sigma_e^2})
\end{equation*}

\textit{Sampling $\sigma_a^2$}
\begin{equation*}
    \textnormal{log}p(\sigma_a^2|.) \propto -(\frac{1}{2} + h_a + 1)\textnormal{log}(\sigma_a^2) - (\frac{\beta_{a}^2}{2}+l_a)\sigma_a^{-2}
\end{equation*}
\begin{equation*}
    p(\sigma_a^2|.) \sim \textnormal{inverse-gamma}(\frac{1}{2} + h_a, \frac{\beta_{a}^2}{2}+l_a)
\end{equation*}

\textit{Sampling $\sigma_e^2$}
\begin{equation*}
    \textnormal{log}p(\sigma_e^2|.) = -(\frac{n}{2} + h_1 + 1)\textnormal{log}(\sigma_e^2) - (\frac{\sum_{i=1}^n (Y_i - \boldsymbol{M_i}^\top\boldsymbol{\beta_m} - A_i\beta_a - \boldsymbol{C_i}^\top\boldsymbol{\beta_c})^2}{2} +l_1)\sigma_e^{-2}
\end{equation*}
\begin{equation*}
    p(\sigma_e^2|.) \sim \textnormal{inverse-gamma}(\frac{n}{2} + h_1, \frac{\sum_{i=1}^n (Y_i - \boldsymbol{M_i}^\top\boldsymbol{\beta_m} - A_i\beta_a - \boldsymbol{C_i}^\top\boldsymbol{\beta_c})^2}{2}+l_1)
\end{equation*}

\textit{Sampling $\sigma_g^2$}
\begin{eqnarray*}
    \textnormal{log}p(\sigma_g^2|.) &=& -(\frac{pn}{2} + h_2 + 1)\textnormal{log}(\sigma_g^2) \\
    && - (\frac{\sum_{i=1}^n (\boldsymbol{M_i}^\top - A_i\boldsymbol{\alpha_a} - \boldsymbol{C_i}^\top\boldsymbol{\alpha_c})(\boldsymbol{M_i}^\top - A_i\boldsymbol{\alpha_a} - \boldsymbol{C_i}^\top\boldsymbol{\alpha_c})^\top}{2} +l_2)\sigma_g^{-2}
\end{eqnarray*}
\begin{equation*}
    p(\sigma_g^2|.) \sim \textnormal{inverse-gamma}(\frac{pn}{2} + h_2, \frac{\sum_{i=1}^n (\boldsymbol{M_i}^\top - A_i\boldsymbol{\alpha_a} - \boldsymbol{C_i}^\top\boldsymbol{\alpha_c})(\boldsymbol{M_i}^\top - A_i\boldsymbol{\alpha_a} - \boldsymbol{C_i}^\top\boldsymbol{\alpha_c})^\top}{2} +l_2)
\end{equation*}

\textit{Sampling $\boldsymbol{\theta}_0, \boldsymbol{\theta}_1$} \\
We update each of the $\theta_{0k}, \theta_{1k}, k \in \{ 1, 2, 3, 4 \}$ using a double Metropolis-Hastings (DMH) algorithm \citep{liang2010double}. For example, for updating $\theta_{0k}$, we first propose a new $\theta^{\star}_{0k}$ from $\textnormal{N}(\theta_{0k}, \tau^2_{\theta})$ and then simulate an auxiliary variable $\boldsymbol{\gamma}^{\star}$ starting from $\boldsymbol{\gamma}$ based on the new $\boldsymbol{\theta}^{\star}_0, \boldsymbol{\theta}_1$ where all
the elements are the same as $\boldsymbol{\theta}_0, \boldsymbol{\theta}_1$, excluding $\theta_{0k}$. The proposed value $\theta^{\star}_{0k}$ will be accepted with probability min(1, $r_{\theta}$) and the Hastings ratio is,
\begin{equation*}
    r_{\theta} = \frac{\phi(\theta^{\star}_{0k};\mu_{0k}, \sigma_{0k}^2)p(\boldsymbol{\gamma}^{\star}|\boldsymbol{\theta}_0, \boldsymbol{\theta}_1)p(\boldsymbol{\gamma}|\theta^{\star}_0, \boldsymbol{\theta}_1)}{\phi(\theta_{0k};\mu_{0k}, \sigma_{0k}^2)p(\boldsymbol{\gamma}|\boldsymbol{\theta}_0, \boldsymbol{\theta}_1)p(\boldsymbol{\gamma}^{\star}|\theta^{\star}_0, \boldsymbol{\theta}_1)}
\end{equation*}
where $\phi(\theta_{0k};\mu_{0k}, \sigma_{0k}^2)$ is the pdf for the normal distribution $\textnormal{N}(\mu_{0k}, \sigma_{0k}^2)$. The form of $p(\boldsymbol{\gamma}|\boldsymbol{\theta}_0, \boldsymbol{\theta}_1)$ is given by Equation (3.1) as in the main text and the normalizing constants are canceled in the ratio.

\textit{Sampling $\beta_{cw}$}
\begin{equation*}
    \textnormal{log}p(\beta_{cw}| .) = - \sum_{i=1}^n \{ \frac{(C_{iw}\beta_{cw})^2}{2\sigma_e^2} + \sigma_e^{-2}C_{iw}(Y_i - \boldsymbol{M_i}^\top\boldsymbol{\beta_m} - A_i\beta_a - \sum_{s \neq w}C_{iw}\beta_{cw})\beta_{cw} \}
\end{equation*}
\begin{equation*}
    p(\beta_{cw}|.) \sim N( \frac{\sum_{i=1}^nC_{iw}(Y_i - A_i\beta_a - \boldsymbol{M_i}^\top\boldsymbol{\beta_m} - \sum_{s \neq w}C_{iw}\beta_{cw})}{\sum_{i=1}^n C_{iw}^2}, \frac{\sigma_e^2}{\sum_{i=1}^n C_{iw}^2})  \\
\end{equation*}

\textit{Sampling $(\boldsymbol{\alpha_{cw}})_j$}
\begin{equation*}
    \textnormal{log}p((\boldsymbol{\alpha_{cw}})_j| .) = - \sum_{i=1}^n \{ \frac{(C_{iw}(\boldsymbol{\alpha_{cw}})_j)^2}{2\sigma_g^2} + \sigma_g^{-2}C_{iw} (M_i^{(j)} - A_i\alpha_{aj} - \sum_{s \neq w}C_{is}(\boldsymbol{\alpha_{cs}})_j)(\boldsymbol{\alpha_{cw}})_j \}
\end{equation*}
\begin{equation*}
    p((\boldsymbol{\alpha_{cw}})_j|.) \sim N(\frac{\sum_{i=1}^nC_{iw} (M_i^{(j)} - A_i\alpha_{aj} - \sum_{s \neq w}C_{is}(\boldsymbol{\alpha_{cs}})_j)}{\sum_{i=1}^n C_{iw}^2}, \frac{\sigma_g^2}{\sum_{i=1}^n C_{iw}^2}) \\
\end{equation*}

\textit{Swendsen-Wang algorithm}\\
We propose to use Swendsen-Wang algorithm \citep{higdon1998auxiliary} to update the Markov random field, $\boldsymbol{\gamma}$. It is a particular case of auxiliary variable methods. In applying SW, we introduce ``bond variables", $\boldsymbol{u} = \{ u_{ij}, i \sim j \}$, for each neighbor pair $i \sim j$. Given $\boldsymbol{\gamma}$, the non-negative random variable $u_{ij}$ are assumed to be independent and uniformly distributed as below,
\begin{eqnarray}
    p(u_{ij}|\boldsymbol{\gamma}) = \textnormal{exp} \{ -\sum_{k=1}^{4} \theta_{1k}I[ \gamma_i = \gamma_j = k ]\} \times I[ 0 \leq u_{ij} \leq \textnormal{exp} \{ \sum_{k=1}^{4} \theta_{1k}I[ \gamma_i = \gamma_j = k ] \}] \nonumber \\
    p(\boldsymbol{u} |\boldsymbol{\gamma}) = \prod_{i \sim j} \textnormal{exp} \{ -\sum_{k=1}^{4} \theta_{1k}I[ \gamma_i = \gamma_j = k ]\} \times I[ 0 \leq u_{ij} \leq \textnormal{exp} \{ \sum_{k=1}^{4} \theta_{1k}I[ \gamma_i = \gamma_j = k ] \}]
\label{eq:c1}
\end{eqnarray}
Furthermore,
\begin{equation}
    p(\boldsymbol{\gamma}|\boldsymbol{u
    }, \cdot) \propto p(\boldsymbol{\beta_m}, \boldsymbol{\alpha_a} | \boldsymbol{\gamma}) \textnormal{exp} \{ \sum_{i=1}^{p} \sum_{k=1}^{4}\theta_{0k} I[ \gamma_i = k ] \} \times \prod_{i \sim j} I[ 0 \leq u_{ij} \leq \textnormal{exp} \{ \sum_{k=1}^{4} \theta_{1k}I[ \gamma_i = \gamma_j = k ] \}]
\label{eq:c2}
\end{equation}
To sample from the joint posterior of $\boldsymbol{\gamma}$ and $\boldsymbol{u} = \{ u_{ij}, i \sim j \}$, we can iteratively sample from Equation (\ref{eq:c1}) and (\ref{eq:c2}). To sample from (\ref{eq:c2}), we note that $u_{ij} > 1$ implies that $\gamma_i = \gamma_j$, so that the bond variable $\boldsymbol{u}$ partitions mediators into same-labeled clusters, and this happens with a probability of $1-\textnormal{exp} \{ -\sum_{k=1}^{4} \theta_{1k}I[ \gamma_i = \gamma_j = k ] \}$. For a particular cluster, $C$, the probability of belonging to component $k$ is $\propto \prod_{i \in C} p((\boldsymbol{\beta_m})_i, (\boldsymbol{\alpha_a})_i |\gamma_i) \textnormal{exp} \{ \theta_{0k} \gamma_i \}$, and each cluster can be updated independently in turn according to its conditional distribution. The SW implementation can be described as below:
\begin{enumerate}
    \item Update each bond variable according to a uniform distribution:
    \begin{equation*}
        u_{ij}|\boldsymbol{\gamma} \sim U[0, \textnormal{exp} \{ \sum_{k=1}^{4} \theta_{1k}I[ \gamma_i = \gamma_j = k ] \} ]
    \end{equation*}
    Bonds are forbidden from forming wherever the two neighbors are in different groups.
    \item Form the same-labeled clusters (connected components) induced by $u_{ij}$
    \begin{enumerate}[label=\roman*.]
        \item The Union-Find algorithm
        \item Simplifies in the 1-D case
    \end{enumerate}
    \item For each cluster $C$, update its label according to its conditional distribution,
    \begin{equation*}
        p(\gamma_C = k|\cdot) \propto \prod_{i \in C} p((\boldsymbol{\beta_m})_i, (\boldsymbol{\alpha_a})_i |\gamma_i) \textnormal{exp} \{ \theta_{0k} \gamma_i \}, k = 1, 2, 3, 4
    \end{equation*}
\end{enumerate}
We alternate between Swendsen-Wang updates of $\boldsymbol{\gamma}$ and single site Gibbs updates to ensure movement in large patches.

\section{Posterior Sampling Algorithm Details for GMM-CorrS}
\hfill \break
To sample from the posterior distribution using the P\'olya-Gamma method, simply iterate two steps:

\textit{Sampling $w_{jk}$ for each $j$ and $k$}
\begin{equation*}
    w_{jk} | \cdot \sim \textnormal{P\'olya-Gamma}(n_{jk}, b_{kj})
\end{equation*}
where $n_{jk} = 1-\sum_{k^{'} < k}I(\gamma_{j} = k^{'})$, $n_{j1} = 1$. The samples from P\'olya-Gamma distribution can be generated using the algorithm and software in \cite{polson2013bayesian}.

\textit{Sampling $\boldsymbol{b_k}$} \\
We can rewrite 4-dimensional multinomial in terms of 3 binomial densities $\widetilde{\pi}_{j1}$, $\widetilde{\pi}_{j2}$ and $\widetilde{\pi}_{j3}$. Specifically,
\begin{eqnarray*}
    p(\boldsymbol{b_k}) &\propto& \prod_{j} \widetilde{\pi}_{j1}^{I(\gamma_{j} = 1)}((1-\widetilde{\pi}_{j1})\widetilde{\pi}_{j2})^{I(\gamma_{j} = 2)}((1-\widetilde{\pi}_{j1})(1-\widetilde{\pi}_{j2})\widetilde{\pi}_{j3})^{I(\gamma_{j} = 3)} \\
    && ((1-\widetilde{\pi}_{j1})(1-\widetilde{\pi}_{j2})(1-\widetilde{\pi}_{j3})\widetilde{\pi}_{j4})^{I(\gamma_{j} = 4)} \textnormal{MVN}(\boldsymbol{a_k}, \sigma_{dk}^2\boldsymbol{D}) \\
    &\propto& \prod_{j} \widetilde{\pi}_{j1}^{I(\gamma_{j} = 1)}(1-\widetilde{\pi}_{j1})^{1-I(\gamma_{j} = 1)}\widetilde{\pi}_{j2}^{I(\gamma_{j} = 2)}(1-\widetilde{\pi}_{j2})^{I(\gamma_{j} = 3) + I(\gamma_{j} = 4)}\widetilde{\pi}_{j3}^{I(\gamma_{j} = 3)}(1-\widetilde{\pi}_{j3})^{I(\gamma_{j} = 4)} \\ && \textnormal{MVN}(\boldsymbol{a_k}, \sigma_{dk}^2\boldsymbol{D}) 
\end{eqnarray*}
\begin{eqnarray*}
    I(\gamma_{j} = 1) &\sim& \textnormal{Binom}(1, \textnormal{expit}(b_{1j})), \\
    I(\gamma_{j} = 2) &\sim& \textnormal{Binom}(n_{j2}, \textnormal{expit}(b_{2j})), \\
    I(\gamma_{j} = 3) &\sim& \textnormal{Binom}(n_{j3}, \textnormal{expit}(b_{3j}))
\end{eqnarray*}
The multinomial distribution is now expressed with three binomial distributions involving $b_{kj}$, k = 1,2,3. Following the derivation in \cite{polson2013bayesian}, we will have,
\begin{equation*}
    \boldsymbol{b_k} | \cdot \sim \textnormal{MVN}(\boldsymbol{\mu_{bk}},  \boldsymbol{V_{bk}})
\end{equation*}
where
\begin{eqnarray*}
    \boldsymbol{V_{bk}} = (\boldsymbol{\Omega} + (\sigma_{dk}^2)^{-1}\boldsymbol{D}^{-1})^{-1} \\
    \boldsymbol{\mu_{bk}} = \boldsymbol{V_{bk}}(\boldsymbol{\kappa_k} + (\sigma_{dk}^2)^{-1}\boldsymbol{D}^{-1}\boldsymbol{a_k})
\end{eqnarray*}
where $\boldsymbol{\Omega}$ is the diagnol matrix of $w_{jk}$'s, and $\boldsymbol{\kappa_k} = (I(\gamma_{1}=k)-n_{1k}/2, I(\gamma_{2}=k)-n_{2k}/2, ..., I(\gamma_{p}=k)-n_{pk}/2)$. Then we can update $\boldsymbol{\pi_j}$ accordingly.

\textit{Sampling $\sigma_{dk}^2$}
\begin{equation*}
    \sigma_{dk}^2 | \cdot \sim \textnormal{IG}(u + \frac{p}{2}, v + \frac{(\boldsymbol{b_k}-\boldsymbol{a_k})^{T}\boldsymbol{D}^{-1}(\boldsymbol{b_k}-\boldsymbol{a_k})}{2})
\end{equation*}

The other parameters can be sampled in a similar way as in the GMM-Potts, with details described in the previous section.

\section{Empirical FDR Results and Computing Time in Simulations} 
\hfill \break
To estimate the FDR and identify a significance threshold for declaring active mediators, we compute the local false discovery rate for each mediator following \cite{efron2007size}. We define the local false discovery rate for the $j$-th mediator being in the active group as $\textnormal{locfdr}_{j1}$, and it can be expressed as $1-P(\gamma_{j} = 1|\textnormal{ Data})$. We first sort $\textnormal{locfdr}_{j1}$ from the smallest to the largest, where the $j$th ordered value is $\textnormal{locfdr}_{1}^{(j)}, j = 1, ..., p$. Then the cutoff value $c_{1}$ for $\textnormal{locfdr}_{j1}$ to guarantee a 10\% FDR can be identified from,
\begin{equation*}
    \argmax_{c_{1}}{\frac{1}{\sum_{j=1}^p I(\textnormal{locfdr}_{1}^{(j)} < c_{1})}\sum_{j=1}^p I(\textnormal{locfdr}_{1}^{(j)} < c_{1})\textnormal{locfdr}_{1}^{(j)} < 0.1}
\end{equation*}
where $I$ is an indicator function. Following \cite{newton2004detecting}, we declare mediators with an $\textnormal{locfdr}_{j1}$ smaller than the threshold $c_{1}$ as active mediators.

As another practical procedure, we also consider a cutoff on the posterior inclusion probabilities (PIP) to declare active mediators. To evaluate the performance of those significance rules, we report the empirical FDR and TPR in Table \ref{tbl:c4efdr1} and \ref{tbl:c4efdr2} under all the simulation scenarios.

\begin{table}[!h]
\begin{adjustwidth}{-0.7cm}{}
\scalebox{0.85}{
\centering{
  \begin{tabular}{c  c c c c c c c}
    \hline     
    Method & TPR & TPR(locfdr) & FDR(locfdr) & TPR(PIP$>$0.5) & FDR(PIP$>$0.5) & TPR(PIP$>$0.9) & FDR(PIP$>$0.9)\\
    \hline
    \multicolumn{8}{c}{$\rho_1 = 0.5 - 0.03|i-j|, \rho_2 = 0$, Signals in one block} \\
    \hline
    GMM-CorrS & 0.78 & 0.69(0.021) & 0.04(0.008) & 0.82(0.019) & 0.12(0.012) & 0.49(0.016) & 0.02(0.006) \\
    GMM-Potts & 0.93 & 0.79(0.019) & 0.05(0.007) & 0.86(0.014) & 0.07(0.010) & 0.61(0.017) & 0.01(0.002) \\
    \hline
    \multicolumn{8}{c}{$\rho_1 = 0.5 - 0.03|i-j|, \rho_2 = 0$, Signals in two blocks} \\
    \hline
    GMM-CorrS & 0.62 & 0.52(0.018) & 0.07(0.010) & 0.67(0.021) & 0.14(0.012) & 0.40(0.015) & 0.01(0.005) \\
    GMM-Potts & 0.49 & 0.34(0.041) & 0.06(0.025) & 0.66(0.023) & 0.22(0.022) & 0.24(0.032) & 0.02(0.017) \\
    \hline
    \multicolumn{8}{c}{$\rho_1 = 0.9 - 0.05|i-j|, \rho_2 = 0.1$, Signals in one block} \\
    \hline
    GMM-CorrS & 0.81 & 0.49(0.020) & 0.06(0.013) & 0.83(0.014) & 0.17(0.007) & 0.36(0.018) & 0.02(0.015) \\
    GMM-Potts & 0.92 & 0.51(0.043) & 0.05(0.015) & 0.83(0.049) & 0.08(0.014) & 0.23(0.014) & 0.01(0.012) \\
	\hline
	\multicolumn{8}{c}{$\rho_1 = 0.9 - 0.05|i-j|, \rho_2 = 0.1$, Signals in two blocks} \\
    \hline
    GMM-CorrS & 0.49 & 0.31(0.032) & 0.09(0.023) & 0.55(0.032) & 0.23(0.023) & 0.22(0.021) & 0.05(0.020) \\
    GMM-Potts & 0.40 & 0.27(0.006) & 0.06(0.005) & 0.43(0.038) & 0.17(0.022) & 0.17(0.006) & 0.03(0.008) \\
    \hline
	\multicolumn{8}{c}{$\rho_1 = 0$, Signals in two blocks} \\
    \hline
    GMM-CorrS & 0.52 & 0.44(0.015) & 0.03(0.008) & 0.50(0.015) & 0.07(0.012) & 0.35(0.012) & 0.01(0.004) \\
    GMM-Potts & 0.46 &  0.42(0.022) & 0.06(0.016) & 0.50(0.016) & 0.19(0.019) & 0.33(0.016) & 0.02(0.011)\\
    \hline
    \multicolumn{8}{c}{Weak correlation from MESA, Signals in two blocks} \\
    \hline
    GMM-CorrS & 0.44 & 0.32(0.009) & 0.03(0.009) & 0.39(0.011) & 0.08(0.013) & 0.27(0.007) & 0.01(0.006) \\
    GMM-Potts & 0.40 & 0.35(0.014) & 0.07(0.015) & 0.45(0.017) & 0.23(0.022) & 0.27(0.010) & 0.03(0.011) \\
    \hline
\end{tabular}}}
\captionof{table}{Empirical estimates of TPR and FDR in simulations of $n=100, p = 200$. The results are based on 200 replicates for each setting, and the standard errors are shown within parentheses. TPR is the true positive rate controlled at a fixed FDR of 10\%; TPR(locfdr) and FDR(locfdr) are the empirical estimates based on the local FDR approach; TPR(PIP$>$0.9) and FDR(PIP$>$0.9) are the empirical estimates when the PIP threshold for identifying active mediators is 0.9; TPR(PIP$>$0.5) and FDR(PIP$>$0.5) are the empirical estimates when the PIP threshold for identifying active mediators is 0.5.}
\label{tbl:c4efdr1}
\end{adjustwidth}
\end{table}

\begin{table}[!h]
\begin{adjustwidth}{-0.7cm}{}
\scalebox{0.85}{
\centering{
  \begin{tabular}{c  c c c c c c c}
    \hline     
    Method & TPR & TPR(locfdr) & FDR(locfdr) & TPR(PIP$>$0.5) & FDR(PIP$>$0.5) & TPR(PIP$>$0.9) & FDR(PIP$>$0.9)\\
    \hline
    \multicolumn{8}{c}{$\rho_1 = 0.5 - 0.02|i-j|, p_{11} = 100$, Signals in five block} \\
    \hline
    GMM-CorrS & 0.92 & 0.90(0.001) & 0.08(0.002) & 0.88(0.002) & 0.02(0.012) & 0.80(0.003) & 0.00(0.001) \\
    GMM-Potts & 0.97 &  0.96(0.002) & 0.09(0.002) & 0.96(0.002) & 0.01(0.002) & 0.93(0.002) & 0.00(0.002) \\
    \hline
    \multicolumn{8}{c}{Weak correlation from MESA, $p_{11} = 100$, Signals in five blocks} \\
    \hline
    GMM-CorrS & 0.83 &  0.83(0.002) & 0.12(0.003) & 0.81(0.003) & 0.04(0.003) & 0.77(0.003) & 0.00(0.001) \\
    GMM-Potts & 0.76 & 0.86(0.017) & 0.33(0.022) & 0.88(0.010) & 0.35(0.024) & 0.82(0.011) & 0.16(0.017) \\
    \hline
    \multicolumn{8}{c}{$\rho_1 = 0.5 - 0.02|i-j|, p_{11} = 10$, Signals in two blocks} \\
    \hline
    GMM-CorrS & 0.83 & 0.82(0.007) & 0.09(0.006) & 0.81(0.007) & 0.05(0.007) & 0.74(0.009) & 0.01(0.003) \\
    GMM-Potts & 0.85 &  0.88(0.007) & 0.34(0.012) & 0.95(0.007) & 0.65(0.008) & 0.83(0.008) & 0.04(0.012) \\
	\hline
	\multicolumn{8}{c}{$\rho_1 = 0.25, p_{11} = 10$, Signals in two blocks} \\
    \hline
    GMM-CorrS & 0.82 & 0.80(0.006) & 0.09(0.007) & 0.80(0.006) & 0.06(0.007) & 0.74(0.007) & 0.01(0.003) \\
    GMM-Potts & 0.61 & 0.79(0.018) & 0.35(0.043) & 0.81(0.011) & 0.56(0.037) & 0.75(0.010) & 0.15(0.047) \\
    \hline
\end{tabular}}}
\captionof{table}{Empirical estimates of TPR and FDR in simulations of $n=1000, p = 2000$, $p_{11}$ is the number of true active mediators. The results are based on 200 replicates for each setting, and the standard errors are shown within parentheses. TPR is the true positive rate controlled at a fixed FDR of 10\%; TPR(locfdr) and FDR(locfdr) are the empirical estimates based on our PIP approach; TPR(PIP$>$0.9) and FDR(PIP$>$0.9) are the empirical estimates when the PIP threshold for identifying active mediators is 0.9; TPR(PIP$>$0.5) and FDR(PIP$>$0.5) are the empirical estimates when the PIP threshold for identifying active mediators is 0.5.}
\label{tbl:c4efdr2}
\end{adjustwidth}
\end{table}

We performed simulations on a single core of Intel(R) Xeon(R) Platinum 8176 CPU @ 2.10GHz, and the runtime comparison of the proposed methods is shown in Table \ref{table:runtime}. For both the small sample scenario with $n = 100$, $p = 200$, and the large sample scenario with $n = 1000$, $p = 2000$, the proposed algorithms can be finished in a reasonable amount of time. We still acknowledge that future development of new algorithms and/or new methods will likely be required to scale our method to handle thousands of individuals and millions of mediators. 

\begin{table}
\centering
{
  \begin{tabular}{c  c  c }
    \hline                  
	  Method & $n = 100$, $p = 200$ & $n = 1000$, $p = 2000$  \\
      \hline
      GMM-CorrS & 3.5 min & 9.8 hr \\
      GMM-Potts &  2.2 min & 4.0 hr \\
      \hline
\end{tabular}}
\caption{The average runtime of the proposed methods for $n = 100$, $p = 200$ and $n = 1000$, $p = 2000$ in the simulations. Comparison was carried out on a single core of Intel(R) Xeon(R) Platinum 8176 CPU @ 2.10GHz. For the proposed methods, we in total ran 150,000 iterations.} 
\label{table:runtime}
\end{table}

\section{Detailed Description of MESA Data}
\hfill \break
MESA is a population-based longitudinal study designed to identify risk factors for the progression of subclinical cardiovascular disease (CVD) \citep{bild2002multi}. A total of 6,814 non-Hispanic white, African-American, Hispanic, and Chinese-American women and men aged 45$-$84 without clinically apparent CVD were recruited between July 2000 and August 2002 from the following 6 regions in the US: Forsyth County, NC; Northern Manhattan and the Bronx, NY; Baltimore City and Baltimore County, MD; St. Paul, MN; Chicago, IL; and Los Angeles County, CA. Each field center recruited from locally available sources, which included lists of residents, lists of dwellings, and telephone exchanges. Neighborhood socioeconomic disadvantage scores for each neighborhood were created based on a principal components analysis of 16 census-tract level variables from the 2000 US Census. These variables reflect dimensions of education, occupation, income and wealth, poverty, employment, and housing. For the neighborhood measures, we use the cumulative average of the measure across all available MESA examinations. The descriptive statistics for the exposure and outcome can be found in Table \ref{tbl:descriptive}.

In the MESA data, between April 2010 and February 2012 (corresponding to MESA Exam 5), DNAm were assessed on a random subsample of 1,264 non-Hispanic white, African-American, and Hispanic MESA participants aged 55$-$94 from the Baltimore, Forsyth County, New York, and St. Paul field centers. After excluding respondents with missing data on one or more variables, we had phenotype and DNAm data from purified monocytes on a total of 1,225 individuals and we focused on this set of individuals for analysis. The detailed description of DNAm data collection, quantitation and data processing procedures can be found in Liu et al \citep{liu2013methylomics}. Briefly, the Illumina HumanMethylation450 BeadChip was used to measure DNAm, and bead-level data were summarized in GenomeStudio. Quantile normalization was performed using the \textit{lumi} package with default settings \citep{du2008lumi}. Quality control (QC) measures included checks for sex and race/ethnicity mismatches and outlier identification by multidimensional scaling plots. Further probe filtering criteria included: ``detected'' DNAm levels in $<$90\% of MESA samples (detection $p$-value cut-off = 0.05), existence of a SNP within 10 base pairs of the target CpG site, overlap with a non-unique region, and suggestions by DMRcate \citep{chen2013discovery} (mostly cross-reactive probes). Those procedures leave us 403,713 autosomal probes for analysis. 

For computational reasons, we first selected a subset of CpG sites to be used in the final multivariate mediation analysis model. In particular, for each single CpG site in turn, we fit the following linear mixed model to test the marginal association between the CpG site and the exposure variable:
\begin{equation}
    M_i = A_i\psi_a + \boldsymbol{C_{1i}}^\top\boldsymbol{\psi_c} + \boldsymbol{Z_i}^\top\boldsymbol{\psi_u} + \epsilon_i, i = 1, ..., n
\end{equation}
where $A_i$ represents neighborhood SES value for the $i$'th individual and $\psi_a$ is its coefficient; $\boldsymbol{C_{1i}}$ is a vector of covariates that include age, gender, race/ethnicity, childhood socioeconomic status, adult socioeconomic status and enrichment scores for each of 4 major blood cell types (neutrophils, B cells, T cells and natural killer cells) to account for potential contamination by non-monocyte cell types; $\boldsymbol{Z_i}^\top\boldsymbol{\psi_u}$ represent methylation chip and position random effects and are used to control for possible batch effects. The error term $\epsilon_i \sim \textnormal{MVN} (0, \sigma^2 I_n)$ and is independent of the random effects. We obtained $p$-values for testing the null hypothesis $\psi_a=0$ from the above model. We further applied the R/Bioconductor package BACON \citep{van2017controlling} to these $p$-values to further adjust for possible inflation using an empirical null distribution. Based on these marginal $p$-values, we selected top 2,000 CpG sites with the smallest $p$-values for our Bayesian multivariate analysis.

\begin{table}[!h]
\centering{
  \begin{tabular}{l c c c}
    \hline                  
	& \begin{tabular}[x]{@{}c@{}}\textbf{Full}\\\textbf{Sample}\\\textbf{(n, \%)}\end{tabular} &  \begin{tabular}[x]{@{}c@{}}\textbf{Neighborhood}\\\textbf{Socioeconomic}\\\textbf{Disadvantage}\\\textbf{Mean (SD)}\end{tabular} & \begin{tabular}[x]{@{}c@{}}\textbf{Glucose}\\\textbf{Mean (SD)}\end{tabular} \\
	\hline
    \textbf{Full sample} & 1225 (100) & -0.32 (1.11) & 29.5 (5.49) \\
    \textbf{Age} & & & \\
    \multicolumn{1}{r}{55$-$65 years} & 462 (38) & -0.18 (0.96) & 30.3 (6.02)   \\
    \multicolumn{1}{r}{66$-$75 years} & 397 (32) & -0.30 (1.16) & 30.1 (5.21) \\
    \multicolumn{1}{r}{76$-$85 years} & 300 (24) & -0.47 (1.15) & 28.2 (4.65)\\
    \multicolumn{1}{r}{86$-$95 years} & 66 (5) & -0.67 (1.46) & 26.6 (4.66) \\
    \textbf{Race/ethnic group} & & & \\
    \multicolumn{1}{r}{Non-Hispanic white} & 580 (47) & -0.56 (1.18) & 28.7 (5.40) \\
    \multicolumn{1}{r}{African-American} & 263 (22) & -0.16 (0.98) & 30.5 (5.69) \\
    \multicolumn{1}{r}{Hispanic} & 382 (31) & -0.05 (1.00) & 30.0 (5.32) \\
    \textbf{Gender} & & & \\
    \multicolumn{1}{r}{Female} & 633 (52) & -0.24 (1.09) & 30.1 (6.20) \\
    \multicolumn{1}{r}{Male} & 592 (48) & -0.40 (1.12) & 28.9 (4.54)  \\
    \hline
\end{tabular}}
\caption{Characteristics of 1225 participants from MESA. \%: proportion in the corresponding category. SD: standard deviation. }
\label{tbl:descriptive}
\end{table}

\section{Detailed Description of LIFECODES Data}
\hfill \break
The LIFECODES prospective birth cohort enrolled approximately 1,600 pregnant women between 2006 and 2008 at the Brigham and Women’s Hospital in Boston, MA. Participants between
20 and 46 years of age were all at less than 15 weeks gestation at the initial study visit, and followed up to four visits (targeted at median 10, 18, 26, and 35 weeks gestation). At the initial study visit, questionnaires were administered to collect demographic and health-related information. Subjects' urine and plasma samples were collected at each study visit. Among participants recruited in the LIFECODES cohort, 1,181 participants were followed until delivery and had live singleton infants. The birth outcome, gestational age, was also recorded at delivery, and preterm birth was defined as delivery prior to 37 weeks gestation. This study received institutional review board (IRB) approval from the Brigham and Women’s Hospital and all participants provided written informed consent. All of the methods within this study were performed in accordance with the relevant guidelines and regulations approved by the IRB. Additional details regarding recruitment and study design can be found in \citep{mcelrath2012longitudinal, ferguson2014variability}. 

In this study, we focused on a subset of $n = 161$ individuals with their urine and plasma samples collected at one study visit occurring between 23.1 and 28.9 weeks gestation (median = 26 weeks). Characteristics of the subset sample is described in Table \ref{tbl:descriptive1}. Subjects' urine samples were refrigerated ($4^\circ$C) for a maximum of 2 hours before being processed and stored at $-80^\circ$C. Approximately 10mL of blood was collected using ethylenediaminetetraacetic acid plasma tubes and temporarily stored at $4^\circ$C for less than 4 hours. Afterwards, blood was centrifuged for 20 minutes and stored at $-80^\circ$C. Environmental exposure analytes were measured from urine samples by NSF International in Ann Arbor, MI, following the methods developed by the Centers for Disease Control (CDC) \citep{silva2007quantification}. Those exposure analytes include phthalates, phenols and parabens, trace metals and polycyclic aromatic hydrocarbons. To adjust for urinary dilution, specific gravity (SG) levels were measured in each urine sample using a digital handheld refractometer (ATAGO Company Ltd., Tokyo, Japan), and was included as a covariate in regression models. Urine and plasma were subsequently analyzed for endogenous biomarkers, including 51 eicosanoids, five oxidative stress biomarkers and five immunological biomarkers in the present study. For a detailed description of the biomarkers that we analyzed and the media (urine or plasma) in which they were measured, please refer to \cite{aung2019prediction}.

\begin{table}[!h]
\centering{
  \begin{tabular}{l c c c}
    \hline                  
	& \begin{tabular}[x]{@{}c@{}}\textbf{Full}\\\textbf{Sample}\\\textbf{(n = 161)}\end{tabular} &  \begin{tabular}[x]{@{}c@{}}\textbf{Preterm}\\\textbf{(<37 weeks gestation,}\\\textbf{n = 52)}\end{tabular} & \begin{tabular}[x]{@{}c@{}}\textbf{Control}\\\textbf{(n = 109)}\end{tabular} \\
	\hline
    \textbf{Age$^\mathrm{a}$} & 32.7 (4.4) & 32.1 (5.0) & 33.0 (4.2) \\
    \textbf{BMI at Initial Visit$^\mathrm{a}$} & 26.7 (6.4) & 28.5 (7.6) & 25.8 (5.6) \\
    \textbf{Race/ethnic group$^\mathrm{b}$} & & & \\
    \multicolumn{1}{r}{White} & 102 (63\%) & 29 (56\%) & 73 (67\%) \\
    \multicolumn{1}{r}{African-American} & 18 (11\%) & 7 (13\%) & 11 (10\%)\\
    \multicolumn{1}{r}{Other} & 41 (26\%) & 16 (31\%) & 25 (23\%) \\
    \textbf{Gestational weeks$^\mathrm{a}$} & 37.5 (3.1) & 34.1 (3.2) & 39.1 (1.1)\\
    \hline
\end{tabular}}
\captionof{table}{Characteristics of all participants in the subset sample from the LIFECODES prospective birth cohort
(n = 161). $^\mathrm{a}$Continuous variables presented as: mean (standard deviation). $^\mathrm{b}$Categorical variables
presented as: count (proportion).}
\label{tbl:descriptive1}
\end{table}

We also utilize the biological pathway information to construct neighbors in GMM-Potts and correlation matrix in GMM-CorrS. The results are shown in Table \ref{tbl:data1}.

\begin{table}[!h]
\centering{
   \begin{tabular}{ cccc }
   \hline
	Method & Selected Mediators & PIP & $\hat{\beta}_{mj}\hat{\alpha}_{aj}$ (95\% CI)\\
    \hline
    \multicolumn{4}{c}{\textit{ Polycyclic aromatic hydrocarbons $\rightarrow$ Biomarkers $\rightarrow$ \textit{Gestational Age}}} \\
    \hline
     GMM-Potts & 8(9)-EET & 0.99 & 0.698(0.391, 1.005) \\
     & 9,10-DiHOME & 0.91 & -0.359(-0.671, 0.000)\\
     \hdashline
     GMM-CorrS & 12(13)-EpoME & 1.00 & 1.132(0.867, 1.426) \\
     & 9-oxoODE & 0.99 & -0.834(-1.119, -0.535) \\
     \hdashline
     GMM  & 8(9)-EET & 1.00 & 0.698(0.407, 0.990) \\
     & 9,10-DiHOME & 0.99 & -0.394(-0.693, -0.091) \\
    \hline
\end{tabular}
}
\caption{Summary of the identified active mediators from the data application on LIFECODES study based on 10\% FDR with the local FDR approach. Both GMM-Potts and GMM-CorrS use the neighborhood structure based on biological pathways. Besides the PIP, we also report the effect estimation $\hat{\beta}_{mj}\hat{\alpha}_{aj}$ and its 95\% credible interval.}
\label{tbl:data1}
\end{table}

\bibliographystyle{ba}
\bibliography{reference}